\documentclass[10pt,draftcls,onecolumn]{IEEEtran}

\usepackage{times,latexsym,amsfonts,amsmath,amssymb,mathrsfs,verbatim,cite}
\usepackage{epsfig,epsf}
\usepackage{graphicx}
\usepackage[dvips]{color}

\usepackage{txfonts}

\def\rR{{\mathbb R}}

\def\QED{\mbox{\rule[0pt]{1.5ex}{1.5ex}}}

\def\@begintheorem#1#2{\tmpitemindent\itemindent\topsep 0pt\rm\trivlist
    \item[\hskip \labelsep{\indent\it #1\ #2:}]\itemindent\tmpitemindent}
\def\@opargbegintheorem#1#2#3{\tmpitemindent\itemindent\topsep 0pt\rm \trivlist
    \item[\hskip\labelsep{\indent\it #1\ #2\
    \rm(#3):}]\itemindent\tmpitemindent}
\def\@endtheorem{\endtrivlist\unskip}

\renewcommand{\theequation}{\arabic{section}.\arabic{equation}}

\begin{document}

\title{On Delayed Sequential Coding of Correlated Sources$^{\text{\small 1}}$}


\author{\authorblockN{Nan Ma and Prakash Ishwar}\\
\authorblockA{Department of Electrical and Computer Engineering \\
Boston University, Boston, MA 02215 \\ {\tt \{nanma, pi\}@bu.edu} }}

\maketitle


\begin{abstract}
Motivated by video coding applications, the problem of sequential
coding of correlated sources with encoding and/or decoding
frame-delays is studied.  The fundamental tradeoffs between
individual frame rates, individual frame distortions, and
encoding/decoding frame-delays are derived in terms of a
single-letter information-theoretic characterization of the
rate-distortion region for general inter-frame source correlations
and certain types of potentially frame specific and coupled
single-letter fidelity criteria. The sum-rate-distortion region is
characterized in terms of generalized directed information measures
highlighting their role in delayed sequential source coding
problems. For video sources which are spatially stationary
memoryless and temporally Gauss--Markov, MSE frame distortions, and
a sum-rate constraint, our results expose the optimality of
idealized differential predictive coding among all causal sequential
coders, when the encoder uses a positive rate to describe each
frame. Somewhat surprisingly, causal sequential encoding with
one-frame-delayed noncausal sequential decoding can \emph{exactly
match} the sum-rate-MSE performance of \emph{joint coding} for all
nontrivial MSE-tuples satisfying certain positive semi-definiteness
conditions. Thus, even a single frame-delay holds potential for
yielding significant performance improvements. Generalizations to
higher order Markov sources are also presented and discussed. A
rate-distortion performance equivalence between, causal sequential
encoding with delayed noncausal sequential decoding, and, delayed
noncausal sequential encoding with causal sequential decoding, is
also established.
\end{abstract}


\section{Introduction}
\addtocounter{footnote}{+1} \footnotetext{This material is based
upon work supported by the US National Science Foundation (NSF)
under award (CAREER) CCF--0546598. Any opinions, findings, and
conclusions or recommendations expressed in this material are those
of the authors and do not necessarily reflect the views of the NSF.
Parts of this work were presented at ITA'07 and ISIT'07.}

 Differential predictive coded modulation (DPCM) is a popular and
well-established sequential predictive source compression method
with a long history of development (see
\cite{Farvardin,Berbook,Ber,Gibson,Ish,Zamir,Rose1,Rose2} and the
references therein). DPCM has had wide impact on the evolution of
compression standards for speech, image, audio, and video coding.
The classical DPCM system consists of a causal sequential predictive
encoder and a causal sequential decoder.  This is aligned with
applications having low delay tolerance at both encoder and decoder.
However, there are many interesting scenarios where these
constraints can be relaxed. There are three additional sequential
source coding systems possible when limited delays are allowed at
the encoder and/or the decoder: (i) causal (C) encoder and noncausal
(NC) decoder; (ii) NC-encoder and C-decoder; and (iii) NC-encoder
and NC-decoder.  Application examples of these include,
respectively, non-real-time display of live video for C--NC,
zero-delay display of non-real-time encoded video for NC--C, and
non-real-time display of non-real-time video for NC--NC (see
Figs.~\ref{fig:vidcod}, \ref{fig:CCarch}, \ref{fig:otherarch} and
\ref{fig:NCNC}). Of special interest, for performance comparison, is
joint coding (JC) which may be interpreted as an extreme special
case of the C--NC, NC--C, and the NC--NC systems where all frames
are jointly processed and jointly reconstructed
(Fig.~\ref{fig:otherarch}(c)).

The goal of this work is to provide a computable (single-letter)
characterization of the fundamental information-theoretic
rate-distortion performance limits for the different scenarios and
to quantify and compare the potential value of systems with limited
encoding and decoding delays in different rate-distortion regimes.
The primary motivational application of our study is video coding
(see Section~\ref{sec:motapps}) with encoding and decoding {\em
frame} delays.\footnote{Accordingly, terms like frame-delay and
``causal'' and ``noncausal'' encoding and/or decoding should be
interpreted within this application context.}

To characterize the fundamental tradeoffs between individual
frame-rates, individual expected frame-distortions, encoding and
decoding frame-delays, and source inter-frame correlation, we build
upon the information-theoretic framework of sequential coding of
correlated sources.  This mathematical framework was first
introduced in \cite{Vis} (and independently studied in
\cite{Tatikonda1,Tatikonda2} under a stochastic control framework
involving dynamic programming) within the context of the {\em purely
C--C}\footnote{The terminology is ours.} (i.e., without
frame-delays). sequential source coding system. As noted in
\cite{Vis}, the results for the well-known successive-refinement
source coding problem (see \cite{Equitz,Rimoldi, Venkataramani}) can
be derived from those for the C--C sequential source coding problem
by setting all sources to be identically equal to the same source.
The complete (single-letter) rate-distortion region for two sources
(with a remark regarding generalization to multiple sources) and
certain types of perceptually-motivated coupled single-letter
distortion criteria were derived in \cite{Vis}. Our results cover
not only the two-frame C--C problem studied in \cite{Vis} but also
the C--NC, the NC--C, the NC--NC, and the JC cases for arbitrary
number of sources and for general coupled single-letter distortion
criteria. We have also been able to simplify some of the key
derivations in \cite{Vis} (the C--C case).

The benefits of decoding delay on the rate versus MSE performance
was investigated in \cite{Ish}, where the video was modeled as a
Gaussian process which is spatially independent and temporally
first-order-\-autoregressive. An idealized DPCM structure was
imposed on both the encoder and the decoder. In contrast to
conventional rate-distortion studies of {\em scalar} DPCM systems
based on {\em scalar quantization} and {\em high-rate} asymptotics
(see \cite{Farvardin,Berbook,Ber} and references therein),
\cite{Ish} studied DPCM systems with vector-valued sources and large
spatial (as opposed to high rate) asymptotics similar in spirit to
\cite{Vis,Tatikonda1,Tatikonda2} but with decoding frame-delays. The
main findings of \cite{Ish} were that (i) NC-decoders offer a
significant {\em relative} improvement in the MSE at medium to low
rates for video sources with strong temporal correlation, (ii) most
of this improvement can be attained with a modest decoding
frame-delay, and (iii) the gains vanish at very high and very low
rates.

In contrast to the insistence on DPCM encoders and decoders in
\cite{Ish}, here we consider arbitrary rate-constrained coding
structures. When specialized to spatially stationary memoryless,
temporally Gauss--Markov video sources, with MSE as the fidelity
metric and a sum-rate constraint, our results reveal the
information-theoretic optimality of idealized DPCM encoders and
decoders for the C--C sequential coding system (Corollary~1.3). A
second, somewhat surprising, finding is that for $k$-th order
Gauss--Markov video sources with a sum-rate constraint, a C-encoder
with a $k$-frame-delayed NC-decoder can {\em exactly match} the
sum-rate-MSE performance of the {\em joint coding system}
 which can wait to collect {\em all}
frames of the video segment before jointly processing and jointly
reconstructing them\footnote{This is similar to the coding of {\em
correlated} parallel vector Gaussian sources but with an {\em
individual} MSE constraint on each source component.}
(Corollary~5.2). Interestingly, this performance equivalence does
not hold for all MSE-tuples. It holds for a {\em non-trivial} subset
which satisfies certain positive semi-definiteness conditions. The
performance-matching region expands with increasing frame-delays
allowed at the decoder until it completely coincides with the set of
all reachable tuples of the JC system. A similar phenomenon holds
for Bernoulli-Markov sources with a Hamming distortion metric. Thus,
the benefit of even a single frame-delay can be significant.
These two specific architectural results constitute the main
contributions of this work.

For clarity of exposition, the proofs of achievability and converse
coding theorems in this paper are limited to discrete, memoryless,
({\em spatially}) stationary (DMS) correlated sources taking values
in finite alphabets and bounded (but coupled) single-letter fidelity
criteria. Analogous results can be established for continuous
alphabets (e.g., Gaussian sources) and unbounded distortion criteria
(e.g., MSE) using the techniques in \cite{OohamaIT98} but are not
discussed here.

The rest of this paper is organized as follows. Delayed sequential
coding systems and their associated operational rate-distortion
regions are formulated in Section~\ref{sec:problem}. To preserve the
underlying intuition and flow of ideas, we first focus on 3-stage
coding systems and then present natural extensions to general
$T$-stage coding systems. Coding theorems and associated
implications for the C--C, JC, C--NC, and NC--C systems are
presented in Sections III, IV, V and VI respectively. Results for
$T$-stage C--NC and NC--NC systems are presented in Sections VII and
VIII. A detailed proof of achievability and converse coding theorems
is presented only for the C--NC system with $T=3$ frames. The
achievability and converse results for other delayed coding systems
are similar but lengthy, repetitive, and cumbersome, and are
therefore omitted. We conclude in Section IX.

{\bf Notation:} The nonnegative cone of real numbers is denoted by
$\rR^+$ and `iid' denotes independent and identically distributed.
Vectors are denoted in boldface (e.g., ${\bf x},~{\bf X}$). The
dimension of the vector will be clear from the context. With the
exception of $T$ denoting the size of a group of pictures (GOP) in a
video segment and $R$ denoting a rate, random quantities are denoted
in upper case (e.g., $X,~{\bf X}$), and their specific
instantiations in lower case (e.g., $X = x,~{\bf X} = {\bf x}$).
When $A$ denotes a random variable, $A^n$ denotes the ordered tuple
$(A_1,\ldots,A_n)$, $A_m^n$ denotes $(A_m,\ldots,A_n)$, and $A(i-)$
denotes $(A(1),\ldots,A(i-1))$. However, for a set $\mathcal A$,
$\mathcal A^n$ denotes the $n$-fold Cartesian product $\mathcal A
\times \ldots \times \mathcal A$. For a function $g(a)$,
$g^n(a(1),\ldots,a(n))$ denotes the samplewise function
$(g(a(1)),\ldots,g(a(n)))$.

\section{\label{sec:problem}Problem formulation}

\subsection{Statistical model for $T$ correlated sources}

$T$ correlated DMSs taking values in finite alphabets are defined by
\[\left(X_1(i),\ldots,X_T(i)\right)_{i=1}^n \in
\left(\mathcal{X}_1 \times \ldots \times \mathcal{X}_T\right)^n,\]
\[|\mathcal X_j| < \infty,\  \forall j=1,\ldots,T.\]
The joint probability distribution of sources is given by
\[\text{for }i=1,\ldots,n,\ \ (X_1(i),\ldots,X_T(i)) \sim \mbox{iid } p_{X_1 \ldots X_T}(x_1,\ldots,x_T).\]
Potentially, the (spatially) iid assumption can be relaxed to
spatially stationary ergodic by a general AEP argument, but is not
treated in this paper. Of interest are the large-$n$ asymptotics of
achievable rate and distortion tuples.

\subsection{Video coding application context \label{sec:motapps}}

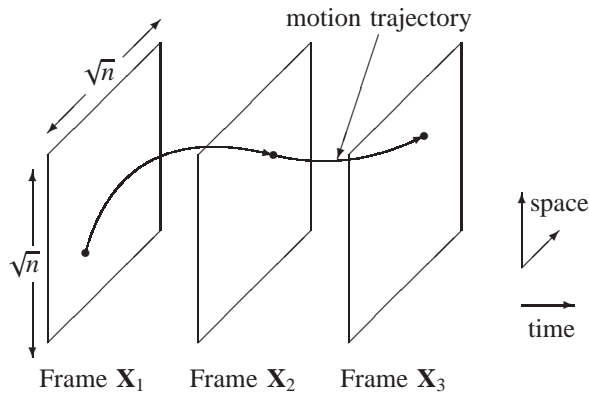
\begin{figure}[!htb]
\centering
\begin{picture}(9,4.5) 
\put(1.2,0.4){
    \put(0,0){  \put(0,0){\line(1,1){1.5}}
                \put(0,2.5){\line(1,1){1.5}}
                \put(0,0){\line(0,1){2.5}}
                \put(1.5,1.5){\line(0,1){2.5}}
                \put(0.6,-0.5){\makebox(0,0){Frame $\mathbf X_1$}}}
    \put(2,0){  \put(0,0){\line(1,1){1.5}}
                \put(0,2.5){\line(1,1){1.5}}
                \put(0,0){\line(0,1){2.5}}
                \put(1.5,1.5){\line(0,1){2.5}}
                \put(0.6,-0.5){\makebox(0,0){Frame $\mathbf X_2$}}}
    \put(4,0){  \put(0,0){\line(1,1){1.5}}
                \put(0,2.5){\line(1,1){1.5}}
                \put(0,0){\line(0,1){2.5}}
                \put(1.5,1.5){\line(0,1){2.5}}
                \put(0.6,-0.5){\makebox(0,0){Frame $\mathbf X_3$}}}
    \put(-0.2,-0.2){\put(0,1){\vector(0,-1){1}}
                    \put(0,1.5){\vector(0,1){1}}
                    \put(-0.15,1.25){\makebox(0,0){$\sqrt{n}$}}}
    \put(0,2.8)    {\put(0.5,0.5){\vector(-1,-1){0.5}}
                    \put(1,1){\vector(1,1){0.5}}
                    \put(0.65,0.8){\makebox(0,0){$\sqrt{n}$}}}
    \qbezier(0.5,1.2)(1,3)(3,2.5)
    \qbezier(3,2.5)(4,2.25)(5,2.75)
    \put(0.5,1.2){\circle*{.1}}
    \put(3,2.5){\circle*{.1}}
    \put(5,2.75){\circle*{.1}}
    \put(2.8,2.55){\vector(4,-1){0.2}}
    \put(4.8,2.65){\vector(2,1){0.2}}
    \put(4.4,4.1){\vector(-1,-3){0.55}}
    \put(4.4,4.3){\makebox(0,0){motion trajectory}}
    \put(6.3,1){
        \put(0,0){\vector(1,1){0.5}}
        \put(0,0){\vector(0,1){1}}
        \put(0,-.5){\vector(1,0){0.7}}
        \put(0.65,0.8){\makebox(0,0){}}
        \put(0.4,-.8){\makebox(0,0){time}}
        \put(0.5,0.8){\makebox(0,0){space}}}
    }

\end{picture}
\caption{\label{fig:vidcod}{\small \sl Illustrating
motion-compensated video coding for $T=3$ frames.} }
\end{figure}

In Fig.~\ref{fig:vidcod}, $\mathbf{X}_1, \ldots, \mathbf{X}_T$
represent $T$ {\em video frames} with ${\bf X}_j = (X_j(i))_{i=1}^n,
j=1,\ldots,T$. Here, $i$ denotes discrete index of the spatial
location of a picture element (pixel) relative to a certain spatial
scan order (e.g., zig-zag or raster scan), and $X_j(i)$ denotes
discrete pixel intensity level at spatial location $i$ in frame
number $j$. Instead of being available simultaneously for encoding,
initially, only $(X_1(i))_{i=1}^n$ is available, then
$(X_2(i))_{i=1}^n$ ``arrives'', followed by $(X_3(i))_{i=1}^n$, and
so on. This temporal structure captures the order in which the
frames are processed. The statistical structure assumed in
Section~II.A above implies that the
sources are {\em spatially independent but temporally dependent}. \\

While this is rarely an accurate statistical model for the {\em
unprocessed} frames of a video segment in a scene (usually
corresponding to the GOP in video coding standards), it is a
reasonable approximation for the evolution of the video
\emph{innovations process} along optical-flow motion trajectories
for groups of adjacent pixels (see \cite{Ish} and references
therein). This model assumes arbitrary temporal correlation but iid
spatial correlation. The statistical law $p_{X_1 \ldots X_T}$ is
assumed to be known here. In practice, this may be learnt from
pre-operational training using clips from video databases used by
video-codec standardization groups such as H.26x and MPEG-x which is
quite similar in spirit to the offline optimization of quantizer
tables in commercial video codecs. Single-letter
information-theoretic coding results need asymptotics along some
problem dimension to exploit some version of the law of large
numbers. Here, the asymptotics are in the {\em spatial dimension}
and is matched to video coding applications where it is quite
typical to have frames of size $n = 352 \times 288 $ pixels at $30$
frames per second (full CIF\footnote{CIF stands for Common
Intermediate Format. Progressively scanned HDTV is typically $n =
1280\times 720 \approx$ one million pixels at $60$ frames per
second.}). It is also fairly common to code video in groups of $T =
15$ pictures.

\subsection{Delayed sequential coding systems}\label{sec:structures}
For clarity of exposition, we start the discussion with the
exemplary case of three frame systems. Systems with an arbitrary
number of frames are studied in sections~\ref{sec:generalresults}
and \ref{sec:NCNC}.

\begin{figure}[!htb]
\centering
\begin{picture}(9,4.2)  
\put(1.2,0.4){
    \put(2,2.5){\framebox(1.,0.7){\textsf{Enc.1}}}
    \put(2,1.2){\framebox(1.,0.7){\textsf{Enc.2}}}
    \put(2,-0.1){\framebox(1.,0.7){\textsf{Enc.3}}}
    \put(4.1,2.5){\framebox(1.,0.7){\textsf{Dec.1}}}
    \put(4.1,1.2){\framebox(1.,0.7){\textsf{Dec.2}}}
    \put(4.1,-0.1){\framebox(1.,0.7){\textsf{Dec.3}}}
    \put(1.4,2.85){\vector(1,0){0.6}}
    \put(3,2.85){\vector(1,0){1.1}}
    \put(5.1,2.85){\vector(1,0){0.4}}
    \put(1.4,1.5){\vector(1,0){0.6}}
    \put(1.75,1.6){\vector(1,0){0.25}}
    \put(3,1.5){\vector(1,0){1.1}}
    \put(3.85,1.6){\vector(1,0){0.25}}
    \put(5.1,1.55){\vector(1,0){0.4}}
    \put(1.4,0.15){\vector(1,0){0.6}}
    \put(1.65,0.25){\vector(1,0){0.35}}
    \put(1.75,0.35){\vector(1,0){0.25}}
    \put(3,0.15){\vector(1,0){1.1}}
    \put(3.75,0.25){\vector(1,0){0.35}}
    \put(3.85,0.35){\vector(1,0){0.25}}
    \put(5.1,0.25){\vector(1,0){0.4}}
    \put(1.75,2.85){\line(0,-1){2.5}}
    \put(3.85,2.85){\line(0,-1){2.5}}
    \put(1.65,1.5){\line(0,-1){1.25}}
    \put(3.75,1.5){\line(0,-1){1.25}}
    \put(1.75,2.85){\circle*{.1}}
    \put(3.85,2.85){\circle*{.1}}
    \put(1.75,1.6){\circle*{.1}}
    \put(3.85,1.6){\circle*{.1}}
    \put(1.65,1.5){\circle*{.1}}
    \put(3.75,1.5){\circle*{.1}}
    \put(0.3,2.85){\makebox(0,0){$\left(X_1(1)... X_1(n)\right)$}}
    \put(0.3,1.5){\makebox(0,0){$\left(X_2(1)... X_2(n)\right)$}}
    \put(0.3,0.15){\makebox(0,0){$\left(X_3(1)... X_3(n)\right)$}}
    \put(6.6,2.85){\makebox(0,0){$\left(\widehat X_1(1)... \widehat X_1(n)\right)$}}
    \put(6.6,1.55){\makebox(0,0){$\left(\widehat X_2(1)... \widehat X_2(n)\right)$}}
    \put(6.6,0.25){\makebox(0,0){$\left(\widehat X_3(1)... \widehat X_3(n)\right)$}}
    \put(3.4,3.15){\makebox(0,0){$R_1$}}
    \put(3.4,1.8){\makebox(0,0){$R_2$}}
    \put(3.4,0.45){\makebox(0,0){$R_3$}}
    \put(3.4,2.84){\makebox(0,0){/}}
    \put(3.4,1.48){\makebox(0,0){/}}
    \put(3.4,0.13){\makebox(0,0){/}}
    \put(0.3,3.6){\makebox(0,0){\textsf{Video Frames}}}
    \put(3.4,3.6){\makebox(0,0){\textsf{Rates}}}
    \put(6.5,3.6){\makebox(0,0){\textsf{Decoded Frames}}}
    \put(-0.9,2.5){\vector(0,-1){2.7}}
    \put(-0.9,-0.4){\makebox(0,0){\textsf{time}}}
    \put(-0.9,2.5){\vector(1,0){2.3}}
    \put(0.3,2.25){\makebox(0,0){\textsf{space}}}
    }
\end{picture}
\caption{\label{fig:CCarch} \small \sl C--C: Causal (zero-delay)
sequential encoding with causal sequential decoding. Sum-rate
$=R_{sum}^{C-C} = R_1 + R_2 + R_3$. }
\end{figure}
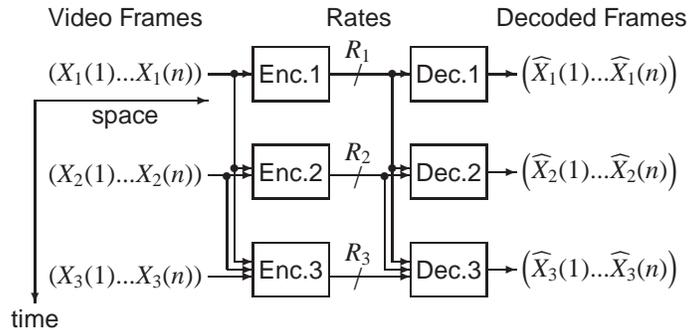

\noindent{$\bullet$}~{\em C--C systems:} The causal (zero-delay)
sequential encoding with (zero-delay) causal sequential decoding
system is illustrated in Fig.~{\ref{fig:CCarch}}. In the first
stage, the video encoder can only access $\mathbf{X}_1$ and encodes
it at rate $R_1$ so that the video decoder is able to reconstruct
$\mathbf{X}_1$ as $\widehat{\mathbf{X}}_1$ immediately. In the
second stage (after one frame-delay), the encoder has access to both
$\mathbf{X}_1$ and $\mathbf{X}_2$ and encodes them at rate $R_2$ so
that the decoder can produce $\widehat{\mathbf{X}}_2$ with help from
the encoder's message in the first stage. In the final stage, the
encoder has access to all the three sources and encodes them at rate
$R_3$ and the decoder produces $\widehat{\mathbf{X}}_3$ with help
from the encoder's messages from all the previous stages. Note that
the processing of information by the video encoder and video decoder
in different stages can be conceptually regarded as distinct source
encoders and source decoders respectively. Also note that it is
assumed that both the encoder and the decoder have enough memory to
store all previous frames and messages.

\begin{figure*}[!htb]
\centering
\begin{picture}(0,5.3)
\put(-9,0){  

 \put(-0.5,1.8){
    \put(2,2.5){\framebox(1.,0.7){\textsf{Enc.1}}}
    \put(2,1.2){\framebox(1.,0.7){\textsf{Enc.2}}}
    \put(2,-0.1){\framebox(1.,0.7){\textsf{Enc.3}}}
    \put(4.1,1.2){\framebox(1.,0.7){\textsf{Dec.1}}}
    \put(4.1,-0.1){\framebox(1.,0.7){\textsf{Dec.2}}}
    \put(4.1,-1.4){\framebox(1.,0.7){\textsf{Dec.3}}}
    \put(1.4,2.85){\vector(1,0){0.6}}
    \put(3,2.85){\line(1,0){0.85}}
    \put(1.4,1.5){\vector(1,0){0.6}}
    \put(1.75,1.6){\vector(1,0){0.25}}
    \put(3,1.5){\vector(1,0){1.1}}
    \put(3.85,1.6){\vector(1,0){0.25}}
    \put(5.1,1.55){\vector(1,0){0.4}}
    \put(1.4,0.15){\vector(1,0){0.6}}
    \put(1.65,0.25){\vector(1,0){0.35}}
    \put(1.75,0.35){\vector(1,0){0.25}}
    \put(3,0.15){\vector(1,0){1.1}}
    \put(3.75,0.25){\vector(1,0){0.35}}
    \put(3.85,0.35){\vector(1,0){0.25}}
    \put(5.1,0.25){\vector(1,0){0.4}}
    \put(3.65,-1.15){\vector(1,0){0.45}}
    \put(3.75,-1.05){\vector(1,0){0.35}}
    \put(3.85,-0.95){\vector(1,0){0.25}}
    \put(5.1,-1.05){\vector(1,0){0.4}}
    \put(1.75,2.85){\line(0,-1){2.5}}
    \put(3.85,2.85){\line(0,-1){3.8}}
    \put(1.65,1.5){\line(0,-1){1.25}}
    \put(3.75,1.5){\line(0,-1){2.55}}
    \put(3.65,0.15){\line(0,-1){1.3}}
    \put(1.75,2.85){\circle*{.1}}
    \put(1.75,1.6){\circle*{.1}}
    \put(3.85,1.6){\circle*{.1}}
    \put(1.65,1.5){\circle*{.1}}
    \put(3.75,1.5){\circle*{.1}}
    \put(3.85,0.35){\circle*{.1}}
    \put(3.75,0.25){\circle*{.1}}
    \put(3.65,0.15){\circle*{.1}}
    \put(1.1,2.85){\makebox(0,0){${\bf X}_1$}}
    \put(1.1,1.5){\makebox(0,0){${\bf X}_2$}}
    \put(1.1,0.15){\makebox(0,0){${\bf X}_3$}}
    \put(5.9,1.55){\makebox(0,0){$\widehat{\bf X}_1$}}
    \put(5.9,0.25){\makebox(0,0){$\widehat{\bf X}_2$}}
    \put(5.9,-1.05){\makebox(0,0){$\widehat{\bf X}_3$}}
    \put(3.4,3.15){\makebox(0,0){$R_1$}}
    \put(3.4,1.8){\makebox(0,0){$R_2$}}
    \put(3.4,0.45){\makebox(0,0){$R_3$}}
    \put(3.4,2.84){\makebox(0,0){/}}
    \put(3.4,1.48){\makebox(0,0){/}}
    \put(3.4,0.13){\makebox(0,0){/}}
    }
\put(5.5,0.5){
    \put(2,2.5){\framebox(1.,0.7){\textsf{Enc.1}}}
    \put(2,1.2){\framebox(1.,0.7){\textsf{Enc.2}}}
    \put(2,-0.1){\framebox(1.,0.7){\textsf{Enc.3}}}
    \put(4.1,2.5){\framebox(1.,0.7){\textsf{Dec.1}}}
    \put(4.1,1.2){\framebox(1.,0.7){\textsf{Dec.2}}}
    \put(4.1,-0.1){\framebox(1.,0.7){\textsf{Dec.3}}}
    \put(1.4,4.15){\line(1,0){0.35}}
    \put(1.75,2.9){\vector(1,0){0.25}}
    \put(1.4,2.8){\vector(1,0){0.6}}
    \put(3,2.85){\vector(1,0){1.1}}
    \put(5.1,2.85){\vector(1,0){0.4}}
    \put(1.4,1.45){\vector(1,0){0.6}}
    \put(1.75,1.65){\vector(1,0){0.25}}
    \put(1.65,1.55){\vector(1,0){0.35}}
    \put(3,1.5){\vector(1,0){1.1}}
    \put(3.85,1.6){\vector(1,0){0.25}}
    \put(5.1,1.55){\vector(1,0){0.4}}
    \put(1.55,0.15){\vector(1,0){0.45}}
    \put(1.65,0.25){\vector(1,0){0.35}}
    \put(1.75,0.35){\vector(1,0){0.25}}
    \put(3,0.15){\vector(1,0){1.1}}
    \put(3.75,0.25){\vector(1,0){0.35}}
    \put(3.85,0.35){\vector(1,0){0.25}}
    \put(5.1,0.25){\vector(1,0){0.4}}
    \put(1.75,4.15){\line(0,-1){3.8}}
    \put(3.85,2.85){\line(0,-1){2.5}}
    \put(1.65,2.8){\line(0,-1){2.55}}
    \put(1.55,1.45){\line(0,-1){1.3}}
    \put(3.75,1.5){\line(0,-1){1.25}}
    \put(1.75,2.9){\circle*{.1}}
    \put(1.65,2.8){\circle*{.1}}
    \put(3.85,2.85){\circle*{.1}}
    \put(1.75,1.65){\circle*{.1}}
    \put(3.85,1.6){\circle*{.1}}
    \put(1.65,1.55){\circle*{.1}}
    \put(1.55,1.45){\circle*{.1}}
    \put(3.75,1.5){\circle*{.1}}
    \put(1.1,4.15){\makebox(0,0){${\bf X}_1$}}
    \put(1.1,2.8){\makebox(0,0){${\bf X}_2$}}
    \put(1.1,1.45){\makebox(0,0){${\bf X}_3$}}
    \put(5.9,2.85){\makebox(0,0){$\widehat{\bf X}_1$}}
    \put(5.9,1.55){\makebox(0,0){$\widehat{\bf X}_2$}}
    \put(5.9,0.25){\makebox(0,0){$\widehat{\bf X}_3$}}
    \put(3.4,3.15){\makebox(0,0){$R_1$}}
    \put(3.4,1.8){\makebox(0,0){$R_2$}}
    \put(3.4,0.45){\makebox(0,0){$R_3$}}
    \put(3.4,2.84){\makebox(0,0){/}}
    \put(3.4,1.48){\makebox(0,0){/}}
    \put(3.4,0.13){\makebox(0,0){/}}
    }
\put(11.5,1.8){
    \put(2,-0.1){\framebox(1.,0.7){\textsf{Enc}}}
    \put(4.1,-0.5){\framebox(1.,1.5){\textsf{Dec}}}
    \put(1.4,2.85){\line(1,0){0.35}}
    \put(1.4,1.5){\line(1,0){0.25}}
    \put(1.4,0.15){\vector(1,0){0.6}}
    \put(1.65,0.25){\vector(1,0){0.35}}
    \put(1.75,0.35){\vector(1,0){0.25}}
    \put(3,0.25){\vector(1,0){1.1}}
    \put(5.1,-0.35){\vector(1,0){0.4}}
    \put(5.1,0.25){\vector(1,0){0.4}}
    \put(5.1,0.85){\vector(1,0){0.4}}
    \put(1.75,2.85){\line(0,-1){2.5}}
    \put(1.65,1.5){\line(0,-1){1.25}}
    \put(1.1,2.85){\makebox(0,0){${\bf X}_1$}}
    \put(1.1,1.5){\makebox(0,0){${\bf X}_2$}}
    \put(1.1,0.15){\makebox(0,0){${\bf X}_3$}}
    \put(5.9,0.85){\makebox(0,0){$\widehat{\bf X}_1$}}
    \put(5.9,0.25){\makebox(0,0){$\widehat{\bf X}_2$}}
    \put(5.9,-0.35){\makebox(0,0){$\widehat{\bf X}_3$}}
    \put(3.4,0.55){\makebox(0,0){$R$}}
    \put(3.4,0.23){\makebox(0,0){/}}
    }
    \put(3,0){\makebox(0,0){(a)}}
    \put(9,0){\makebox(0,0){(b)}}
    \put(15,0){\makebox(0,0){(c)}}
    }
\end{picture}
\caption{\label{fig:otherarch} \small \sl (a) C--NC: Causal
sequential encoding with one-stage delayed noncausal sequential
decoding; (b) NC--C: one-stage delayed noncausal sequential encoding
with causal sequential decoding; (c) JC: $(T-1)$-stage delayed joint
(noncausal) encoding with joint (noncausal) decoding. }
\end{figure*}
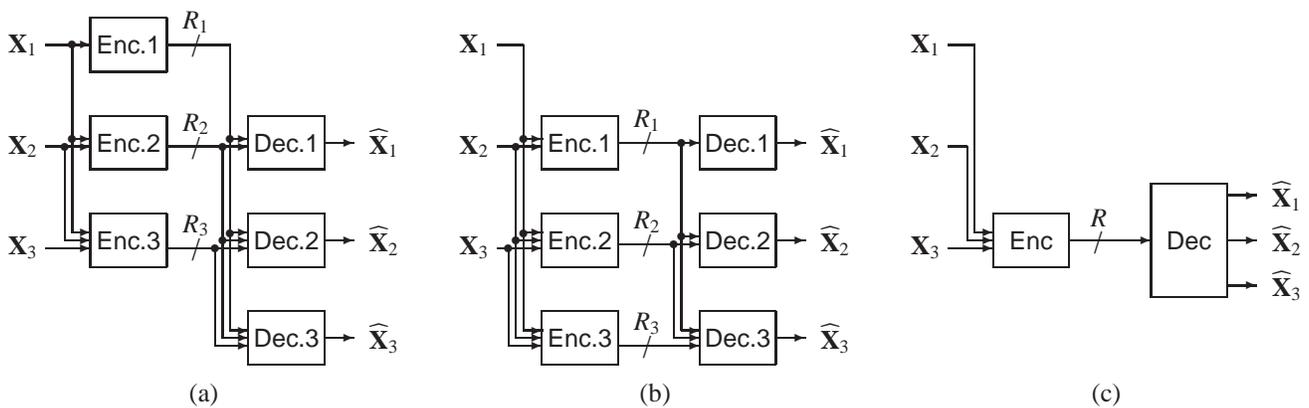

\noindent{$\bullet$}~{\em C--NC systems:} The causal sequential
encoding with one-stage delayed noncausal sequential decoding system
is illustrated in Fig.~\ref{fig:otherarch}(a). In the figure, all
the encoders have access to the same sets of sources as in the C-C
system
 shown in Fig.~\ref{fig:CCarch}. However, the decoders are delayed
(moved downwards) by one stage with respect to
Fig.~\ref{fig:CCarch}. Specifically, the first decoder observes the
messages from the first two encoders to produce
$\widehat{\mathbf{X}}_1$. The second decoder produces
$\widehat{\mathbf{X}}_2$ based on all the three messages from the
three encoders. The third decoder also produces
$\widehat{\mathbf{X}}_3$ using all the messages.
\\
\noindent{$\bullet$}~{\em NC--C systems:} The one-stage delayed
noncausal sequential encoding with causal sequential decoding system
is illustrated in Fig.~\ref{fig:otherarch}(b). Compared with the
C--NC system, the delay
 is on the encoding side. Specifically, the first encoder
 has access to both $\mathbf X_1$ and $\mathbf X_2$. Both the second and the
 third encoder have access to all three sources. The decoders have access to
  the same sets of messages
 sent by the encoders as in the C--C system.\\
\noindent{$\bullet$}~{\em JC systems:} Of special interest is the
{\em joint} (noncausal) encoding and decoding system illustrated in
Fig.~\ref{fig:otherarch}(c). All the sources are collected by a
single encoder and encoded jointly. The single decoder reconstructs
all the frames simultaneously. Note that here the encoding frame
delay is $(T-1)$.

$T$-stage sequential coding systems with $k$-stage frame-delays (see
Fig.~\ref{fig:generalCNC} and \ref{fig:NCNC}) are natural
generalizations of the $3$-stage systems discussed so far. The
general cases will be discussed in detail in
Sections~\ref{sec:generalresults} and \ref{sec:NCNC}.

The C--C blocklength-$n$ encoders and decoders are formally defined by
the maps
\begin{eqnarray*}
(\mbox{Enc.}j) &&f_j^{(n)}:\ \mathcal X_1^{n} \times \ldots \times
\mathcal X_j^{n} \rightarrow
\{1,\ldots,M_j\},\\
(\mbox{Dec.}j) &&g_j^{(n)}:\ \{1,\ldots,M_1\}\times \ldots\times
\{1,\ldots,M_j\} \rightarrow \widehat {\mathcal{X}}_j^{n}
\end{eqnarray*}
for $j=1,\ldots,T$, where $(\log_2 M_j)/n$ is the $j$-th frame coding
rate in bits per pixel (bpp) and $\widehat {\mathcal X}_j$ is the
$j$-th (finite cardinality) reproduction alphabet.

The formal definitions of C--NC encoders are identical to that for
the C--C encoders. However, the C--NC decoders with a $k$-stage
frame-delay are formally defined by the maps
\[
(\mbox{Dec.}j) \ \ g_j^{(n)}:\ \{1,\ldots,M_1\}\times \ldots\times
\{1,\ldots,M_{\min \{j+k,T\}}\} \rightarrow \widehat
{\mathcal{X}}_j^{n},
\]
for $j = 1,\ldots,T$. Similarly, the NC--C decoder definitions are
identical to those for the C--C decoders and the NC--C encoders with
a $k$-stage frame-delay are formally defined by the maps
\[
(\mbox{Enc.}j) \ \ f_j^{(n)}:\ \mathcal X_1^{n} \times \ldots \times
\mathcal X_{\min \{j+k,T\}}^{n} \rightarrow \{1,\ldots,M_j\},
\]
for $j = 1,\ldots,T$. Finally the JC encoder and decoder are defined
by the maps
\begin{eqnarray*}
(\mbox{Enc.}) &&f^{(n)}:\ \mathcal X_1^{n} \times \ldots \times
\mathcal X_T^{n} \rightarrow
\{1,\ldots,M\},\\
(\mbox{Dec.}) &&g^{(n)}:\ \{1,\ldots,M\} \rightarrow \widehat
{\mathcal{X}}_1^{n}\times\ldots\times\widehat {\mathcal{X}}_T^{n}.
\end{eqnarray*}


For a frame-delay $k$, there are boundary effects associated with
the decoders (resp.~encoders) of the last $(k+1)$ frames for the
C--NC (resp.~NC--C) systems. For example, the last two decoders in
Fig.~\ref{fig:otherarch}(a) are operationally equivalent to a single
decoder since both use the same set of encoded messages.
Although redundant, we retain the distinction of the boundary
encoders/decoders for clarity and to aid comparison (see Theorem~4
in Section~\ref{sec:NCC} and Corollary~6.1 in
Section~\ref{sec:NCNC}).

\subsection{Operational rate-distortion regions}
For each $j=1,\ldots,T$, the pixel reproduction quality is measured
by a single-letter distortion criterion. We allow coupled distortion
criteria where the distortion for the current frame can depend on
the reproductions in previous frames:
\[d_j:\mathcal{X}_j \times \widehat{\mathcal{X}}_1 \times \cdots
\times \widehat{\mathcal{X}}_j \rightarrow \rR^+.\] The distortion
criteria are assumed to be bounded, i.e.,
\[d_{j,\max} := \max_{x_j,\hat x_1,\ldots, \hat x_j} d_j(x_j,\hat x_1,\ldots, \hat x_j)< \infty.\]
 The frame
reproduction quality is in terms of the average pixel distortion
\[d_j^{(n)}(\mathbf x_j,\hat{\mathbf x}_1, \ldots, \hat{\mathbf x}_j)=\frac{1}{n}\sum_{i=1}^{n}
d_j(x_j(i), \hat{x}_1(i), \ldots, \hat{x}_j(i)).\] Of interest are
the expected frame distortions $E[d_j^{(n)}( {\bf{X}}_j , \widehat{
\bf{X} }^j)]$.
It is important to notice that these are {\em frame-specific}
distortions as opposed to an average distortion across all frames.
This makes the JC problem different from a standard parallel vector
source coding problem. Also notice that these fidelity criteria
reflect dependencies on previous frame reproductions. For example,
the second distortion criterion is given by $d_2: \mathcal{X}_2
\times \widehat{\mathcal{X}}_1 \times \widehat{\mathcal{X}}_2
\rightarrow \rR^+$, as opposed to a criterion like $\tilde d_2:
\mathcal{X}_2 \times
 \widehat{\mathcal{X}}_2 \rightarrow
\rR^+$ which is independent of previous reproductions. This model is
motivated by the temporal perceptual characteristics of the human
visual system where the visibility threshold at a given pixel
location depends on the luminance intensity of the same pixel in the
previous frames\cite{Vis}.

A rate-distortion-tuple $(\mathbf R, \mathbf D) =
(R_1,\ldots,R_T,D_1,\ldots,D_T)$ is said to be
admissible for a given delayed sequential coding system if, for
every $\epsilon > 0$, and all sufficiently large $n$, there exist
block encoders and decoders satisfying
\begin{eqnarray}
&& \frac{1}{n}\log M_j \leq R_j + \epsilon, \label{eqn:admissible1}\\
&& E[d_j^{(n)}( {\bf{X}}_j ,\widehat{ \bf{X} }^j)] \leq
D_j+\epsilon, \label{eqn:admissible2}
\end{eqnarray}
simultaneously for all $j=1,\ldots,T$. For system $A \in
\{\mbox{C--C, JC}\}$, the operational rate-distortion region
$\mathcal{R}^A$ is the set of all admissible rate-distortion-tuples.
For system $A \in \{\mbox{C--NC, NC--C}\}$ with $k$-stage frame-
delay, the operational rate-distortion region, denoted by
$\mathcal{R}^A_k$, is the set of all admissible
rate-distortion-tuples. We will abbreviate $\mathcal{R}^A_k$ to
$\mathcal{R}^A$ when $k=1$.
The sum-rate region denoted by $\mathcal R_{sum}^A({\bf D})$ (or
$\mathcal R_{k,sum}^A({\bf D})$) is the set of all the admissible
sum-rates $\sum_{j=1}^T R_j$ at the distortion tuple ${\bf D}$.

Note that for any given distortion-tuple the minimum rate of the JC
system is also the minimum sum-rate of a C--NC or NC--C system with
frame-delay $(T-1)$ for the same distortion tuple. For example, in a
$(T-1)$-delayed C--NC system, all the decoders become joint decoders
and the rate-tuple $(R_1 = 0, \ldots, R_{T-1}=0, R_T =
R_{JC}(\mathbf D),\mathbf D)$ is admissible. Hence
$R_{(T-1),sum}^{C-NC}(\mathbf D) = R^{JC}(\mathbf D)$. Therefore
C--NC and NC--C systems for $T=2$ are less interesting. The first
non-trivial delayed sequential coding system arises for $T=3$ (also
see the paragraph after Corollary~5.1). This is the reason for
commencing the discussion with $3$-stage systems.

\section{Results for the 3-stage C--C system}
\subsection{Rate-distortion region}
The C--C rate-distortion region can be formulated as a single-letter
mutual information optimization problem subject to distortion
constraints and natural Markov chains involving auxiliary and
reproduction random
variables and deterministic functions. This characterization is provided by Theorem~1.\\

\noindent{\bf Theorem~1}~{\em (C--C rate-distortion region)} The
single-letter rate-distortion region for a $T=3$ frame C--C system is
given by
\begin{eqnarray}\label{eqn:CCrateregion}
\mathcal{R}^{C-C} &=&
\{(\mathbf R, \mathbf D)~|~\exists~\ U^2, \widehat X^3, g_1(\cdot),g_2(\cdot,\cdot),s.t. \nonumber\\
&& R_1 \geq I(X_1;U_1), \nonumber\\
&& R_2 \geq I(X^2;U_2|U_1), \nonumber\\
&& R_3 \geq I(X^3;\widehat X_3|U^2), \nonumber\\
&& D_j \geq E[d_j(X_j,\widehat{X }^j)], \ \ j=1,2,3, \nonumber\\
&& \widehat X_1 = g_1(U_1),\ \widehat X_2 = g_2(U_1,U_2),\nonumber\\
&& U_1 - X_1 - X_2^3,\  U_2 - (X^2,U_1) - X_3\}
\end{eqnarray}
where $\{U_1, U_2, \widehat X_1, \widehat X_2, \widehat X_3\}$ are
auxiliary and reproduction random variables taking values in
alphabets $\{\mathcal U_1, \mathcal U_2, \widehat{\mathcal{X}}_1,
\widehat{\mathcal{X}}_2, \widehat{\mathcal{X}}_3\}$ satisfying the
cardinality bounds
\begin{eqnarray*}
|\mathcal U_1| & \leq & |\mathcal X_1| + 6, \\
|\mathcal U_2| & \leq & |\mathcal X_1|^2 |\mathcal X_2| + 6|\mathcal
X_1| |\mathcal X_2| +4,
\end{eqnarray*}
and $\{g_1(\cdot),g_2(\cdot,\cdot)\}$ are deterministic functions.
\\

The rate-distortion region in \cite{Vis} is for the 2-stage C--C
problem, whereas the above region is for the 3-stage C--C problem.
The above region differs from what one might expect to get from a
natural extension of the 2-stage  C--C rate-distortion region in
\cite{Vis}. This is because the characterization in Theorem~1 has
different rate inequalities and fewer Markov chain conditions than
what one might expect from the extension. One of the advantages of
the characterization of the rate-distortion region in
(\ref{eqn:CCrateregion}) is that it is more intuitive (as explained
below) and this intuition carries over with little effort to the
case of multiple frames (see Section~VII and VIII). Another
advantage of the characterization of the rate-distortion in
(\ref{eqn:CCrateregion}) is that it is convex and closed as defined.
The convexity can be shown along the lines of the time-sharing
argument in Appendix~\ref{appsec:timesharing} which is part of the
converse proof of the coding theorem for C--NC systems. The
closedness can be shown along the lines of the convergence argument
in Appendix~\ref{appsec:closure}. Therefore, unlike the
characterization provided in \cite{Vis}, there is no need to take
the convex hull and closure in (\ref{eqn:CCrateregion}).

The proof of achievability can be carried out using standard random
coding and random binning arguments and will be similar in spirit to
the derivation for the $T=2$ frame case in \cite{Vis}, but with a
different intuitive interpretation. Hence we will only present the
intuition and informally sketch the steps leading to the proof of
Theorem~1 in the following paragraph. As remarked in the
introduction, a detailed proof of achievability and converse results
will be presented only for the C--NC system with $T=3$ frames
(Appendices~\ref{app:forwardproofthm3} and \ref{app:proofthm3}). The
proofs of achievability and converse results for other systems can
be carried out in a similar manner but the derivations become
lengthy, repetitive, and cumbersome, and are therefore omitted.

The region in Theorem~1 has the following natural interpretation.
First, $\mathbf X_1$ is quantized to $\mathbf U_1$ using a random
codebook-1 for encoder-1 without access to $\mathbf X_2^3$.
Decoder-1 recovers $\mathbf U_1$ and reproduces $\mathbf X_1$ as
$\widehat{\mathbf X}_1=g_1^n(\mathbf U_1)$. Next, the tuple $ \{
\mathbf X^2, \mathbf U_1 \} $ is (jointly) quantized to $\mathbf
U_2$ without access to $\mathbf X_3$ using a random codebook-2 for
encoder-2. The codewords are further randomly distributed into bins
and the bin index of $\mathbf U_2$ is sent to the decoder. Decoder-2
identifies $\mathbf U_2$ from the bin with the help of $\mathbf U_1$
as side-information (available from decoder-1) and reproduces
$\mathbf X_2$ as $\widehat{\mathbf X}_2 = g_2^n(\mathbf U_1,\mathbf
U_2)$. Finally, encoder-3 (jointly) quantizes $\{ \mathbf X^3,
\mathbf U^2 \} $ into $\widehat{\mathbf X}_3$ using encoder-3's
random codebook, bins the codewords and sends the bin index of
$\widehat{\mathbf X}_3$ such that decoder-3 can identify
$\widehat{\mathbf X}_3$ with the help of $\mathbf U^2$ as
side-information available from decoders 1 and 2. The constraints on
the rates and Markov chains ensure that with high probability (for
all large enough $n$) both encoding (quantization) and decoding
(recovery) succeed and the recovered words are jointly strongly
typical with the source words to meet the target distortions. Notice
that the conditioning random variables that appear in the
conditional mutual information expressions at each stage correspond
to quantities that are known to both the encoding and decoding sides
at that stage due to the previous stages. Using this observation,
one can intuitively write down an achievable rate-distortion region
for general delayed sequential coding systems by inspection.

\subsection{Sum-rate region}
The sum-rate region can be obtained from the rate-distortion region
$\mathcal R^{C-C}$ as shown in the following corollary. The main
simplification is the {\em absence} of the auxiliary random
variables $U^2$.

\noindent{\bf Corollary~1.1}~{\em (C--C Sum-rate region)} The
sum-rate region for the C--C system is $\mathcal R_{sum}^{C-C}({\bf
D})=[ R^{C-C}_{sum}({\bf D}), \infty )$ where the minimum sum-rate
is
\begin{equation}\label{eqn:CCsumrate}
R^{C-C}_{sum}({\bf D}) = \min_{  \scriptstyle
  E[d_j(X_j,\widehat{ X }^j)]\leq D_j, j = 1,2,3, \atop
  {  \scriptstyle  \widehat X_1 - X_1 - X_2^3, \ \widehat X_2 - (X^2,\widehat X_1) - X_3}
 } I(X^3;\widehat X^3). \\
\end{equation}

\begin{proof}
For any point $(\mathbf R, \mathbf D)\in \mathcal R^{C-C}$, there
exist auxiliary random variables and functions satisfying all the
constraints in (\ref{eqn:CCrateregion}). Since the Markov chains
$U_1 - X_1 -X_2^3$ and $U_2 - (X^2,U_1) -X_3$ hold, and $\widehat
X^2$ is a function of $U^2$, we have
\begin{eqnarray*}
R_1+R_2+R_3 &\geq & I(X_1;U_1)+I(X^2;U_2|U_1)+I(X^3;\widehat
X_3|U^2)\\
&=& I(X^3;U_1)+I(X^3;U_2|U_1)+I(X^3;\widehat X_3|U^2)
\\
&=& I(X^3;U^2,\widehat X_3)\\
&=& I(X^3;U^2,\widehat X^3)\\
&\geq & I(X^3;\widehat X^3).
\end{eqnarray*}
It can be verified that Markov chains $\widehat X_1 - X_1 - X_2^3$
and $\widehat X_2 - (X^2,\widehat X_1) - X_3$ hold. Therefore the
right hand side of (\ref{eqn:CCsumrate}) is not greater than the
minimum sum rate.

On the other hand, because $\{U_1=\widehat X_1, U_2=\widehat X_2\}$
is a possible choice of $\{U_1, U_2\}$,
\[R^{C-C}_{sum}(\mathbf{D})= \min I(X^3;U^2,\widehat X_3) \leq \min
I(X^3;\widehat X^3),\] where the first minimization is subject to
the constraints in (\ref{eqn:CCrateregion}), and the second
minimization is subject to the constraints in (\ref{eqn:CCsumrate}).
Therefore (\ref{eqn:CCsumrate}) holds.
\end{proof}

As will become clear in the sequel, the minimum sum-rate for any
type of delayed sequential coding system is given by the
minimization of the mutual information between the source random
variables $X^T$ and the reproduction random variables
$\widehat{X}^T$ subject to several expected distortion and
Markov-chain constraints involving these random variables of a form
similar to (\ref{eqn:CCsumrate}).

\subsection{Sum-rate region for Gaussian source and MSE}
In the case of Gaussian sources and MSE distortion criteria, the
minimum sum-rate of any delayed sequential coding system (see
Corollaries 1.1, 3.1, 5.1 and Theorem~2) can be achieved by
reproduction random variables which are jointly Gaussian with the
source random
variables. This is contained in the following lemma.\\

\noindent{\bf Lemma} If $(X_1,\ldots,X_T)$ are jointly Gaussian, the
minimum value of $I(X^T;\widehat X^T)$ subject to MSE constraints
$E[(X_j-\widehat X_j)^2]\leq D_j, j = 1, \ldots, T$ and Markov chain
constraints involving $X^T$ and $\widehat{X}^T$ is achieved by
reproduction random variables $\widehat X^T$ which are jointly
Gaussian with $X^T$. \\

\begin{proof}
Given any reproduction random vector $\widehat{\mathbf X}=(\widehat
X_1, \ldots, \widehat X_T)$ satisfying the MSE and Markov chain
constraints, we can construct a new random vector
$\widetilde{\mathbf X}=(\widetilde X_1, \ldots, \widetilde X_T)$
which is {\em jointly Gaussian} with $\mathbf X = (X_1, \ldots,
X_T)$ with the same second-order statistics. Specifically,
$cov(\widehat{\mathbf X})=cov(\widetilde{\mathbf X})$ and
$cov(\mathbf X, \widehat{\mathbf X})=cov(\mathbf X,
\widetilde{\mathbf X})$. Since MSEs are fully determined from
second-order statistics, $\widetilde{\mathbf X}$ automatically
satisfies the same MSE constraints as $\widehat{\mathbf X}$. The
Markov chain constraints for $\widehat{\mathbf X}$ imply
corresponding conditional uncorrelatedness constraints for
$\widehat{\mathbf X}$, which will also hold for $\widetilde{\mathbf
X}$. Since $\widetilde{\mathbf X}$ is jointly Gaussian, conditional
uncorrelatedness is equivalent to conditional independence.
Therefore $\widetilde{\mathbf X}$ will also satisfy the
corresponding Markov chain constraints.

Let the linear MMSE estimate of $\mathbf X$ based on
$\widehat{\mathbf X}$ be given by $A \widehat {\mathbf X}$ where $A$
is a matrix. Note that by the orthogonality principle and the joint
Gaussianity of ${\bf X}$ and $\widetilde{\bf X}$ we have $ ({\mathbf
X}-A \widehat{\mathbf X}) \perp \widehat{\mathbf X}$, and further $
({\mathbf X}-A \widetilde{\mathbf X}) \Perp \widetilde{\mathbf X}$.
Therefore,
\begin{eqnarray*}
I(\mathbf X; \widehat{\mathbf X}) &=& h({\mathbf X}) - h({\mathbf X}-A \widehat{\mathbf X}|\widehat{\mathbf X})\\
& \geq &h({\mathbf X}) - h({\mathbf X}-A \widehat{\mathbf X})\\
& \stackrel{(b)}{\geq} &h({\mathbf X}) - h({\mathbf X}-A \widetilde{\mathbf X})\\
& \stackrel{(c)}{=} &h({\mathbf X}) - h({\mathbf X}-A
\widetilde{\mathbf X}|\widetilde{\mathbf X})\\ &=& I(\mathbf X;
\widetilde{\mathbf X}).
\end{eqnarray*}
Step (b) is because $( {\mathbf X}-A \widetilde{\mathbf X})$ has the
same second-order statistics as $ ({\mathbf X}-A \widehat{\mathbf
X})$ and it is a {\em jointly Gaussian} random vector. Step (c) is
because $({\mathbf X}-A \widetilde{\mathbf X})$ is independent of
$\widetilde{\mathbf X}$.

In conclusion, given an {\em arbitrary} reproduction vector, we can
construct a {\em Gaussian} random vector $\widetilde{\mathbf X}$
satisfying the same MSE and Markov chain constraints as
$\widehat{\mathbf X}$ and $I(\mathbf X; \widehat{\mathbf X}) \geq
I(\mathbf X; \widetilde{\mathbf X})$. Hence the minimum value of
$I(X^T;\widehat X^T)$ subject to MSE and Markov chain constraints
will be achieved by a reproduction random vector which is jointly
Gaussian with ${\bf X}$.
\end{proof}

Since Gaussian vectors are characterized by means and covariance
matrices, the minimum sum-rate computation reduces to a determinant
optimization problem involving Markov chain and second-order moment
constraints.

For Gauss--Markov sources, $p_{X_1 X_2 X_3}=\mathcal N ({\bf 0},
\Sigma_X)(x_1,x_2,x_3)$ where the covariance matrix $\Sigma_X$ has
the following structure
\begin{equation*}
\Sigma_{X} = \left(
  \begin{array}{ccc}
    \sigma_1^2 & \rho_1 \sigma_1 \sigma_2 & \rho_1 \rho_2 \sigma_1 \sigma_3 \\
    \rho_1 \sigma_1 \sigma_2 & \sigma_2^2 & \rho_2 \sigma_2 \sigma_3 \\
    \rho_1 \rho_2 \sigma_1 \sigma_3 & \rho_2 \sigma_2 \sigma_3 & \sigma_3^2 \\
  \end{array}
\right),
\end{equation*}
which is consistent with the Markov chain relation $X_1-X_2-X_3$
associated with the Gauss--Markov assumption. Define a distortion
region $\mathcal D^{C-C}:=\{{\bf D}~|~D_1 \leq \sigma_1^2, D_2 \leq
\sigma_{W_2}^2, D_3 \leq \sigma_{W_3}^2\}$ where
\begin{eqnarray}\label{eqn:prederror}
\sigma_{W_j}^2=\rho_{j-1}^2
\frac{\sigma_{j}^2}{\sigma_{j-1}^2}D_{j-1}+
(1-\rho_{j-1}^2)\sigma_{j}^2,\ \ j=2,3
\end{eqnarray}
whose significance will be discussed below. The C--C minimum sum-rate
evaluated for any MSE tuple ${\bf D}$ in this region is given by the
following corollary. \\

\noindent{\bf Corollary~1.2}~{\em (C--C minimum sum-rate for
Gauss--Markov sources and MSE)} In the distortion region $\mathcal
D^{C-C}$, the C--C minimum sum-rate for Gauss--Markov sources and MSE
is
\begin{equation}\label{eqn:CCGaussian}
R^{C-CGM}_{sum}(\mathbf{D}) = \frac{1}{2}\log
\left(\frac{\sigma_1^2}{D_1}\right) + \frac{1}{2}\log\left(
\frac{\sigma_{W_2}^2}{D_2}\right)+ \frac{1}{2}\log
\left(\frac{\sigma_{W_3}^2}{D_3}\right).
\end{equation}
\newline

The proof of Corollary~1.2 is given in
Appendix~\ref{app:proofcor12}. The form of (\ref{eqn:CCGaussian})
suggests the following idealized (achievable) coding scheme which is
explained with reference to Fig.~\ref{fig:DPCM} and the upper bound
argument in the proof of Corollary 1.2 in
Appendix~\ref{app:proofcor12}. Encoder-1 initially quantizes
$\mathbf X_1$ into $\widehat{\mathbf{X}}_1$ to meet the target MSE
$D_1$ using an ideal Gaussian rate-distortion quantizer and
decoder-1 recovers $\widehat{\mathbf{X}}_1$. Since the quantizer is
ideal, the joint distribution of $(\mathbf
X_1,\widehat{\mathbf{X}}_1)$ will follow the test-channel
distribution of the rate-distortion function for a memoryless
Gaussian source \cite[p.~345, 370]{CoverThomas}. This idealization
holds in the limit as the blocklength $n$ tends to infinity. Let
$\mathbf W_1:=\mathbf X_1$ and $\widehat {\mathbf
W}_1:=\widehat{\mathbf X}_1$. Next, encoder-2 makes the causal
minimum mean squared error (MMSE) prediction of $\mathbf X_2$ based
on $\widehat{\mathbf{X}}_1$ and quantizes the prediction error
$\mathbf W_2$ into $\widehat {\mathbf W}_2$ using an ideal Gaussian
rate-distortion quantizer so that decoder-2 can form $\widehat{\bf
X}_2$ to meet the target MSE $D_2$ with help from $\widehat {\mathbf
W}_1$. The asymptotic per-component variance of $\mathbf W_2$ will
be consistent with (\ref{eqn:prederror}) because the rate-distortion
quantizer is ideal. Specifically, decoder-2 recovers $\widehat
{\mathbf W}_2$ and creates the reproduction $\widehat{\bf X}_2$ as
the causal MMSE estimate of $\mathbf X_2$ based on $ \widehat
{\mathbf W}^2$. Finally, encoder-3 makes the causal MMSE prediction
of $\mathbf X_3$ based on $\widehat{\mathbf{W}}^2$ and quantizes the
prediction error $\mathbf W_3$ into $\widehat {\mathbf W}_3$ using
an ideal Gaussian rate-distortion quantizer so that decoder-3 can
form $ \widehat{\mathbf{X}}_3$ to meet the target MSE $D_3$ with
help from $\widehat {\mathbf W}^3$. Decoder-3 recovers $\widehat
{\mathbf W}_3$ and makes the reproduction $\widehat{\bf X}_3$ as the
MMSE estimate of $\mathbf{X}_3$ based on $\widehat {\mathbf W}^3 $.
The C--C coding scheme just described is an idealized version of
DPCM (see \cite{Farvardin,Berbook,Ber,Ish,Zamir} and references
therein) because the rate-distortion quantizer is idealized. The
above arguments lead to the following corollary.\\


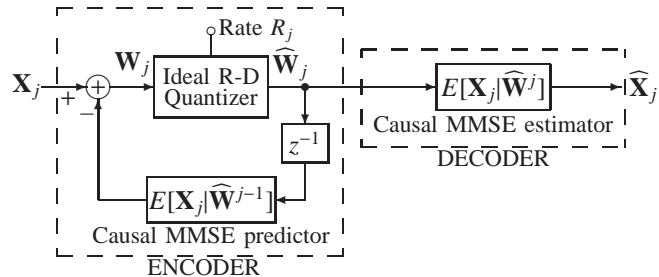
\begin{figure}[!htb]
\centering
\begin{picture}(9,3.5) 
\put(0.75,0.5){
    \put(1.5,1.6){\framebox(1.5,0.8){$\stackrel{\text{\small Ideal R-D}}{\text{\small Quantizer}}$}}
    \put(2.25,2.4){\line(0,1){0.3}}
    \put(2.25,2.75){\circle{0.1}}
    \put(3.2,1.0){\framebox(0.6,0.5){$z^{-1}$}}
    \put(1.4,0.2){\framebox(1.7,0.6){$E[{\bf X}_j|\widehat {\bf W}^{j-1}]$}}
    \put(0.75,2){\circle{.3}}
    \put(0.75,2){\makebox(0,0){+}}
    \put(0.35,1.85){\makebox(0,0){+}}
    \put(0.6,1.6){\makebox(0,0){--}}
    \put(0.2,-0.25){\dashbox{0.2}(3.8,3.3){}}
    \put(5.25,1.7){\framebox(1.5,0.6){$E[\mathbf{X}_j|\widehat {\bf W}^{j}]$}}
    \put(4.25,1.2){\dashbox{0.2}(3.5,1.3){}}
    \put(0,2){\vector(1,0){0.6}}
    \put(0.9,2){\vector(1,0){0.6}}
    \put(3,2){\vector(1,0){2.25}}
    \put(3.5,2){\vector(0,-1){0.5}}
    \put(3.5,1){\line(0,-1){0.5}}
    \put(3.5,0.5){\vector(-1,0){0.4}}
    \put(1.4,0.5){\line(-1,0){0.65}}
    \put(0.75,0.5){\vector(0,1){1.35}}
    \put(3.5,2){\circle*{.1}}
    \put(6.75,2){\vector(1,0){0.95}}
    \put(-0.2,2){\makebox(0,0){$\mathbf X_j$}}
    \put(8,2){\makebox(0,0){$\widehat {\bf X}_j$}}
    \put(1.2,2.3){\makebox(0,0){$\mathbf W_j$}}
    \put(3.3,2.3){\makebox(0,0){${\widehat{\bf W}}_j$}}
    \put(2.25,0){\makebox(0,0){\small Causal MMSE predictor}}
    \put(6,1.5){\makebox(0,0){\small Causal MMSE estimator}}
    \put(2.15,-0.4){\makebox(0,0){\small ENCODER}}
    \put(6,1.05){\makebox(0,0){\small DECODER}}
    \put(2.85,2.75){\makebox(0,0){\small Rate $R_j$}}
    }
\end{picture}
\caption{\label{fig:DPCM} \small \sl Illustrating idealized DPCM.}
\end{figure}

\noindent{\bf Corollary~1.3}~{\em (C--C Optimality of idealized DPCM
for Gauss--Markov sources and MSE)} The C--C minimum sum-rate-MSE
performance for Gauss--Markov sources is achieved by idealized DPCM
for all distortion tuples ${\bf D}$ in the distortion region
$\mathcal
D^{C-C}$. \\

The distortion region $\mathcal D^{C-C}$ is the set of distortion
tuples for which the DPCM encoder uses a positive rate for each
frame. Note that $\mathcal D^{C-C}$ has a
{\em non-zero volume} for nonsingular sources ($\sigma_j\neq 0, \rho_j \neq \pm 1$). 
Hence, the assertion that DPCM is optimal for
C--C systems is a nontrivial statement. \\

\section{Results for the 3-stage JC system}\label{sec:JC}
\noindent{\bf Theorem~2}~{\em (JC rate-distortion function,
\cite[Problem~14, p.134]{Csi})} The single-letter rate-distortion
function for the joint coding system is given by
\begin{equation}\label{eqn:JCrate}
R^{JC}({\bf D}) = \min_{E[d_j(X_j,\widehat{ X }^j)]\leq D_j, \ j =
1,2,3
 } I(X^3;\widehat X^3).
\end{equation}
\newline

Compared to $R^{C-C}_{sum}({\bf D})$ given by (\ref{eqn:CCsumrate}),
the JC rate-distortion function $R^{JC}({\bf D})$ given by
(\ref{eqn:JCrate}) having no Markov chain constraints is a lower
bound for $R^{C-C}_{sum}({\bf D})$. While this follows from a direct
comparison of the single-letter rate-distortion functions, from the
operational structure of C--C, C--NC, NC--C, and JC systems it is
clear that the JC rate-distortion function is in fact a lower bound
for the sum-rates for {\em all} delayed sequential coding systems.

Similar to Corollary~1.2 which is for a C--C system, Gaussian
sources, and MSE
distortion criteria, we have the following corollary for a JC system. \\

\noindent{\bf Corollary~2.1}~{\em (JC rate-MSE function for
  Gauss--Markov sources)}\\
\noindent{(i)} For the distortion region $\mathcal D^{JC}
:=\{\mathbf{D}~|~(\Sigma_X-\mbox{diag}({\bf D}))\geq 0\}$, the JC
rate-MSE function for jointly Gaussian sources is given by
\begin{equation}\label{eqn:JCGauss}
R^{JCGM}({\bf D}) =\frac{1}{2} \log
\left(\frac{ |\Sigma_X|}{D_1 D_2 D_3}\right).
\end{equation}

\noindent{(ii)} For the distortion region $\mathcal D^{JC}$, the JC
rate-MSE function for Gauss--Markov sources is given by
\begin{eqnarray}\label{eqn:JCGaussMarkov}
R^{JCGM}({\bf D}) & = & \frac{1}{2}\log
\left(\frac{\sigma_1^2}{D_1}\right)+\frac{1}{2}\log
\left(\frac{\sigma_2^2 (1-\rho_1^2)}{D_2}\right)+ \nonumber \\
& & \mbox{} + \frac{1}{2}\log \left(\frac{\sigma_3^2 (1-\rho_2^2)}{D_3}\right).
\end{eqnarray}
\newline

Formula (\ref{eqn:JCGauss}) is the Shannon lower bound
\cite{Ber,Berbook} of the JC rate-distortion function. It can be
achieved in the distortion region $\mathcal D^{JC}$ by the test
channel
\begin{equation}
\label{eqn:JCtestchannel} \widehat {\bf X}+\mathbf{Z}=\mathbf{X}
\end{equation}
where $\mathbf{Z}=(Z_1, Z_2, Z_3)$ and $\widehat {\bf X}=(\widehat
X_1, \widehat X_2, \widehat X_3)$ are independent Gaussian vectors
with covariance matrices
\[
\Sigma_Z=\mbox{diag}(\mathbf{D}), \ \  \Sigma_{\widehat
X}=\Sigma_X-\mbox{diag}(\mathbf{D}),
\]
and $\mathbf X=(X_1, X_2, X_3)$. The existence of this channel is
guaranteed by the definition of $\mathcal D^{JC}$. \\

Comparing (\ref{eqn:CCGaussian}) and (\ref{eqn:JCGaussMarkov}) for
${\bf D} \in \mathcal{D}^{JC} \cap \mathcal{D}^{C-C}$ which
generally has a nonempty interior, we find that in general the C--C
sum-rate $R^{C-CGM}_{sum}({\bf D})$ is {\em strictly} greater than
the JC rate $R^{JCGM}({\bf D})$. However, as $\mathbf D \rightarrow
\mathbf 0$, the two rates are asymptotically equal.

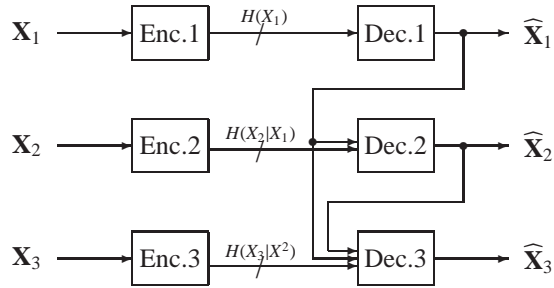
\begin{figure}[htp]
\centering
\begin{picture}(9,3.7) 
\put(0.5,0){
    \put(2,2.9){\framebox(1.,0.7){Enc.1}}
    \put(2,1.4){\framebox(1.,0.7){Enc.2}}
    \put(2,-0.1){\framebox(1.,0.7){Enc.3}}
    \put(5,2.9){\framebox(1.,0.7){Dec.1}}
    \put(5,1.4){\framebox(1.,0.7){Dec.2}}
    \put(5,-0.1){\framebox(1.,0.7){Dec.3}}
    \put(1,3.25){\vector(1,0){1}}
    \put(3,3.25){\vector(1,0){2}}
    \put(6,3.25){\vector(1,0){1}}
    \put(1,1.75){\vector(1,0){1}}
    \put(3,1.7){\vector(1,0){2}}
    \put(4.4,1.8){\vector(1,0){0.6}}
    \put(6,1.75){\vector(1,0){1}}
    \put(1,0.25){\vector(1,0){1}}
    \put(3,0.15){\vector(1,0){2}}
    \put(4.4,0.25){\vector(1,0){0.6}}
    \put(4.6,0.35){\vector(1,0){0.4}}
    \put(6,0.25){\vector(1,0){1}}
    \put(4.6,1){\line(0,-1){0.65}}
    \put(4.6,1){\line(1,0){1.8}}
    \put(6.4,3.25){\line(0,-1){0.75}}
    \put(4.4,2.5){\line(0,-1){2.25}}
    \put(4.4,2.5){\line(1,0){2}}
    \put(6.4,1.75){\line(0,-1){0.75}}
    \put(4.4,1.8){\circle*{.1}}
    \put(6.4,3.25){\circle*{.1}}
    \put(6.4,1.75){\circle*{.1}}
    \put(0.6,3.25){\makebox(0,0){$\mathbf X_1$}}
    \put(0.6,1.75){\makebox(0,0){$\mathbf X_2$}}
    \put(0.6,0.25){\makebox(0,0){$\mathbf X_3$}}
    \put(7.4,3.25){\makebox(0,0){$\widehat{\mathbf X}_1$}}
    \put(7.4,1.75){\makebox(0,0){$\widehat{\mathbf X}_2$}}
    \put(7.4,0.25){\makebox(0,0){$\widehat{\mathbf X}_3$}}
    \put(3.75,3.45){\makebox(0,0){$\scriptstyle H(X_1)$}}
    \put(3.7,1.9){\makebox(0,0){$\scriptstyle H(X_2|X_1)$}}
    \put(3.7,0.35){\makebox(0,0){$\scriptstyle H(X_3|X^2)$}}
    \put(3.7,3.24){\makebox(0,0){/}}
    \put(3.7,1.68){\makebox(0,0){/}}
    \put(3.7,0.13){\makebox(0,0){/}}
    }
\end{picture}
  \caption{Slepian-Wolf coding with a sequential
  decoding
  }
  \label{fig:zerodist}
\end{figure}

We would like to draw some parallels between C--C sequential coding
of correlated  sources and Slepian-Wolf distributed coding of
correlated sources\cite{CoverThomas}. In the Slepian-Wolf coding
problem we have spatially correlated sources, temporal asymptotics,
and a distributed coding constraint. In the C--C sequential coding
problem we have temporally correlated sources, spatial asymptotics,
and a sequential coding constraint. The roles of time and space are
approximately exchanged. In Slepian-Wolf coding, the sources
$\mathbf X_1$, $\mathbf X_2$, and $\mathbf X_3$ can be individually
encoded at the rates $H(X_1)$, $H(X_2|X_1)$, and $H(X_3|X^2)$
respectively and decoded sequentially by first reconstructing
$\mathbf X_1$, then $\mathbf X_2$, and finally $\mathbf X_3$ (see
Fig.~\ref{fig:zerodist}). The sum-rate is equal to the joint entropy
of the three sources which is the rate required for jointly coding
the three sources. The fact that as $\mathbf D \rightarrow \mathbf
0$ the C--C sum-rate approaches the JC sum-rate is consistent with
the fact that in the Slepain-Wolf coding problem, sequential
encoding and decoding does not entail a rate-loss with respect to
joint coding. As $\mathbf D \rightarrow \mathbf 0$ we are
approaching near-lossless compression.

%

\section{Results for the 3-stage C--NC system}
Similar to Theorem~1 and Corollary~1.1 for the C--C system, the
rate-distortion and
sum-rate regions for a C-NC system are characterized by Theorem~3 and Corollary~3.1 respectively as follows. \\

\noindent{\bf Theorem~3}~{\em (C--NC rate-distortion region)} The
single-letter rate-distortion region for a C--NC system with
one-stage decoding frame-delay is given by
\begin{eqnarray}\label{eqn:CNCrateregion}
\mathcal{R}^{C-NC}&=&
\{(\mathbf R, \mathbf D)~|~\exists \ U^2, \widehat X^3, g_1(\cdot,\cdot),s.t.\nonumber\\
&& R_1 \geq I(X_1;U_1), \nonumber\\
&& R_2 \geq I(X^2;U_2|U_1), \nonumber\\
&& R_3 \geq I(X^3;\widehat X_2^3|U^2), \nonumber\\
&& D_j \geq E[d_j(X_j,\widehat{ X }^j)], \ \ j=1,2,3, \nonumber\\
&& \widehat X_1 = g_1(U_1,U_2),\nonumber\\
&& U_1 - X_1 - X_2^3,\  U_2 - (X^2,U_1) - X_3\}
\end{eqnarray}
where $g_1(\cdot,\cdot)$ is a deterministic function and $\{U_1,
U_2\}$ are auxiliary random variables satisfying cardinality bounds
\begin{eqnarray*}
|\mathcal U_1| & \leq & |\mathcal X_1| + 6, \\
|\mathcal U_2| & \leq & |\mathcal X_1|^2 |\mathcal X_2| + 6|\mathcal
X_1| |\mathcal X_2| +5.
\end{eqnarray*}

Note that $ \mathcal R ^{C-C} \subseteq \mathcal R^{C-NC}$ because
the encoders and decoders of a C--C system can also be used in a
C--NC system. As in Theorem~1, the characterization of the
rate-distortion region given in Theorem~3 is both convex and closed
and there is no need to take the convex hull and closure.

The proof of the forward part of Theorem~3 is given in
Appendix~\ref{app:forwardproofthm3}. The region in Theorem~3 has the
following natural interpretation. First, $\mathbf X_1$ is quantized
to $\mathbf U_1$ using a random codebook-1 for encoder-1 without
access to $\mathbf X_2^3$. Next, the tuple $ \{ \mathbf X^2, \mathbf
U_1 \} $ is (jointly) quantized to $\mathbf U_2$ without access to
$\mathbf X_3$ using a random codebook-2 for encoder-2. The codewords
are further randomly distributed into bins and the bin index of
$\mathbf U_2$ is sent to the decoder. Decoder-1 recovers $\mathbf
U_1$ from the message sent by encoder-1. Then it identifies $\mathbf
U_2$ from the bin with the help of $\mathbf U_1$ as side-information
and reproduces $\mathbf X_1$ as $\widehat{\mathbf X}_1 =
g_1^n(\mathbf U_1,\mathbf U_2)$. Finally, encoder-3 (jointly)
quantizes $\{ \mathbf X^3, \mathbf U^2 \} $ into $\widehat{\mathbf
X_2^3}$ using encoder-3's random codebook, bins the codewords and
sends the bin index such that decoder-2 and decoder-3 can identify
$\widehat{\mathbf X_2^3}$ with the help of $\mathbf U^2$ as
side-information available from decoders 1 and 2. The constraints on
the rates and the Markov chains ensure that with high probability
(for all large enough $n$) both encoding (quantization) and decoding
(recovery) succeed and the recovered words are jointly strongly
typical with the source words to meet the target distortions.

The (weak) converse part of Theorem~3 is proved in
Appendix~\ref{app:proofthm3} using standard information inequalities
by defining auxiliary random variables $U_j(i)=(S_j,X_j(i-)), j =
1,2$, where $S_j$ denotes the message sent by the $j$-th encoder
satisfying all the Markov-chain and distortion constraints, and a
convexification (time-sharing) argument as in
\cite[p.397]{CoverThomas}. The cardinality bounds of the auxiliary
random variables are also derived in
Appendix~\ref{appsec:cardinality}
using the Carath\'eodory theorem.\\

\noindent{\bf Corollary~3.1}~{\em (C--NC sum-rate region)} The
sum-rate region for the one-stage delayed C--NC system is $\mathcal
R_{sum}^{C-NC}({\bf D})=[ R^{C-NC}_{sum}({\bf D}),\infty )$ where
the minimum sum-rate is
\begin{equation}\label{eqn:CNCsumrate}
R^{C-NC}_{sum}({\bf D}) = \min_{  \scriptstyle
  E[d_j(X_j,\widehat{ X }^j)] \leq D_j, j = 1,2,3, \atop
  \scriptstyle \widehat X_1 - X^2 - X_3
 } I(X^3;\widehat X^3).
\end{equation}
\newline

\begin{proof}
The proof is similar to that of Corollary~1.1. The main
simplification is the absence of the auxiliary random variables
$U^2$. For any point $(\mathbf R, \mathbf D)\in \mathcal R^{C-NC}$,
there exist auxiliary random variables and functions satisfying all
the constraints in $\mathcal R^{C-NC}$. Since the Markov chains $U_1
- X_1 -X_2^3$ and $U_2 - (X^2,U_1) -X_3$ hold, and $\widehat X_1$ is
a function of $U^2$, we have
\begin{eqnarray*}
R_1+R_2+R_3 &\geq & I(X_1;U_1)+I(X^2;U_2|U_1)+I(X^3;\widehat
X_2^3|U^2)\\
&=& I(X^3;U_1)+I(X^3;U_2|U_1)+I(X^3;\widehat X_2^3|U^2)
\\
&=& I(X^3;U^2,\widehat X_2^3)\\
&=& I(X^3;U^2,\widehat X^3)\\
&\geq & I(X^3;\widehat X^3).
\end{eqnarray*}
It can be verified that the Markov chain $\widehat X_1 - X^2 - X_3$
 holds. Therefore the
right hand side of (\ref{eqn:CNCsumrate}) is not greater than the
minimum sum rate.

On the other hand, because $\{U_1=0, U_2=\widehat X_1\}$ is a
possible choice of $\{U_1, U_2\}$,
\[R^{C-NC}_{sum}(\mathbf{D})= \min I(X^3;U^2,\widehat X_2^3) \leq \min
I(X^3;\widehat X^3),\] where the first minimization is subject to
the constraints in (\ref{eqn:CNCrateregion}), and the second
minimization is subject to the constraints in
(\ref{eqn:CNCsumrate}). Therefore (\ref{eqn:CNCsumrate}) holds.
\end{proof}

As noted earlier, the JC rate-distortion function (\ref{eqn:JCrate})
having no Markov chain constraints is a lower bound for
$R^{C-NC}_{sum}({\bf D})$. Remarkably, for Gauss--Markov sources and
certain nontrivial MSE tuples ${\bf D}$ discussed below,
$R^{C-NC}_{sum}({\bf D})$ coincides with the JC rate
$R^{JC}({\bf D})$.\\

\noindent{\bf Corollary~3.2}~{\em (JC-optimality of a one-stage
delayed C--NC system for Gauss--Markov sources and MSE)} For all
distortion tuples ${\bf D}$ belonging to the distortion region
$\mathcal D^{JC}$ defined in Section~\ref{sec:JC}, Corollary~2.1(i),
we have
\[
R^{C-NCGM}_{sum}({\bf D})=R^{JCGM}({\bf D}).
\]

\begin{proof}
The JC rate-distortion function is achieved by the test channel
(\ref{eqn:JCtestchannel}) in the distortion region $\mathcal
D^{JC}$. We will verify that the Markov chain $\widehat X_1 -
X^2-X_3$ holds for this test channel.

Note that because all the variables are jointly Gaussian, they have
the property that $A \Perp B$ and $A \Perp C$ implies $A \Perp
\{B,C\}$ for any Gaussian vector $(A, B, C)$.

By the Markov chain $X_1 -X_2 - X_3$, the MMSE estimate of $X_3$
based on $X_1$ and $X_2$ is
\begin{equation}\label{appeqn:CNCGaussian}
X_3 = \rho_2 \frac{\sigma_3}{\sigma_2} X_2 + N
\end{equation}
where $N$ is Gaussian and independent of $\{X_1,X_2\}$.

By the structure of the test channel, $Z_1 \Perp \{Z_2, Z_3,
\widehat X_2, \widehat X_3\}$ implies $Z_1 \Perp \{X_2 , X_3\}$,
which further implies $Z_1 \Perp N$. Moreover, because $N \Perp
\{X_1,Z_1\}$, we have $N \Perp \widehat X_1$. Therefore $N \Perp
\{X_1,X_2, \widehat X_1\}$. So the best estimate of $X_3$ based on
$\{X_1,X_2,\widehat X_1\}$ is still formula
(\ref{appeqn:CNCGaussian}). It follows that the Markov chain
$X_3-X_2-(X_1,\widehat X_1)$ holds which in turn implies that
$\widehat X_1 - X^2-X_3$ holds and completes the proof.
\end{proof}

Recall that the JC rate-distortion function is a lower bound for the
minimum sum-rate for all delayed sequential coding systems.
Corollary~3.2 implies that the JC rate-distortion performance is
achievable in terms of sum-rate with only a single frame decoding
delay for Gauss--Markov sources and MSE tuples in the region
$\mathcal D^{JC}$. The first-order Markov assumption on sources
$X_1-X_2-X_3$ is essential for this optimality. An interpretation is
that ${\bf X}_2$ supplies all the help from ${\bf X}_3$ to generate
the optimum $\widehat {\bf X}_1$. More generally (for $T > 3$), as
shown in Section~VII, C--NC encoders need access to only the present
and past frames together with {\em one} future frame to match the
rate-distortion function of the JC system in which {\em all} future
frames are simultaneously available for encoding. Thus, the
neighboring future frame supplies all the help from the entire
future through the Markovian property of sources. The benefit of one
frame-delay is so significant that it is equivalent to arbitrary
frame-delay for Gauss-Markov sources and MSE criteria when $\mathbf
D\in \mathcal D^{JC}$.


It is of interest to compare Corollary~3.2 with the real-time source
coding problem in \cite{Witsenhausen}. In \cite{Witsenhausen} it is
shown that for Markov sources, a C--C encoder may ignore the
previous sources and only use the current source and decoder's
memory without loss of performance. This is a purely structural
result (no spatial asymptotics and computable single-letter
information-theoretic characterizations) exclusively focused on C--C
systems.  In contrast, Corollary~3.2 is about achieving the
JC-system performance with a C--NC system. Additionally,
\cite{Witsenhausen} deals with a frame-averaged expected distortion
criterion as opposed to frame-specific individual distortion
constraints treated here.

The JC-optimality of the one-stage delayed C--NC system is
guaranteed to hold within the distortion region $\mathcal D^{JC}$
defined as the set of all distortion tuples ${\bf D}$ satisfying the
positive semidefiniteness condition $(\Sigma_X - \mbox{diag}({\bf
D})) \geq 0$. For nonsingular sources $\Sigma_X > 0 \Rightarrow
\lambda_{\min}(\Sigma_X) > 0$ where $\lambda_{\min}(\Sigma_X)$ is
the smallest eigenvalue of the positive definite symmetric
(covariance) matrix $\Sigma_X$. For any point $\mathbf D$ in the
closed hypercube $[0,\lambda_{\min}]^T$,
\[\Sigma_X - \mbox{diag}(\mathbf D) = (\Sigma_X - \lambda_{\min} I)+
\mbox{diag}(\lambda_{\min}\mathbf e - \mathbf D)\] where $I$ is the
identity matrix and $\mathbf e=(1,\ldots,1)$ is the all-one vector.
Because both terms are positive semidefinite matrices, the sum is
also positive semidefinite. Therefore $\mathcal D^{JC}$ contains
this hypercube,
 which has a strictly positive
volume in $\rR^T \Rightarrow \mathcal D^{JC}$ has a {\em non-zero
volume}. Hence, the JC-optimality of a C-NC system with one-stage
decoding delay discussed here is a nontrivial assertion. $\mathcal
D^{JC}$ includes all distortion tuples with components below certain
thresholds corresponding to ``sufficiently good'' reproduction
qualities. However, it should be noted that this is {\em not} a
high-rate (vanishing distortion) asymptotic.

On the contrary, the JC-optimality of a C--NC system with one-stage
decoding frame-delay does not hold for all distortion tuples as the
following counter example shows.

 \noindent{\em Counter example:}
Consider Gauss--Markov sources $X^3$ where $X_1=X_2$ and MSE tuple
${\bf D}$ where $D_1=D_2=D$.  The JC problem reduces to a {\em
two-stage JC problem} where the encoder jointly quantizes $(\mathbf
X_1, \mathbf X_3)$ into $(\widehat{\mathbf X}_1, \widehat{\mathbf
X}_3)$ and the decoder simply sets $\widehat{\mathbf X}_2 =
\widehat{\mathbf X}_1$. However, the C--NC problem reduces to a {\em
two-stage C--C problem} with sources $(\mathbf X_1, \mathbf X_3)$
because the first two C--NC encoders are operationally equivalent to
the first C--C encoder observing $ \mathbf X_1 $ and the last C--NC
encoder is operationally equivalent to the second C--C encoder
observing all sources. As mentioned in the last but one paragraph of
Section~\ref{sec:JC}, generally speaking, a two-stage C--C system
does not match (in sum-rate) the JC-system rate-distortion
performance. Therefore the three-stage C--NC system also does not
match the JC performance for these specific sources and certain
distortion tuples ${\bf D}$. Note that these sources are actually
singular ($\Sigma_X$ has a zero eigenvalue) and $\mathcal D^{JC}$
only contains trivial points (either $D = 0$ or $D_3 = 0$). So for
the nontrivial distortion tuples ${\bf D}$ described above (which do
not belong to $\mathcal D^{JC}$), the JC-optimality of a C--NC
system with a one-stage decoding delay fails to hold.

To construct a counter example with nonsingular sources, one can
slightly perturb $\Sigma_X$ such that it becomes positive definite.
However, the JC rate and C--NC sum-rate only change by limited
amounts due to continuity properties of the sum-rate-distortion
function with respect to the source distributions (similar to
\cite[Lemma~2.2, p.124]{Csi}). Therefore we can find a small enough
perturbation such that the rates do not match.

The JC-optimality of the one-stage delayed C--NC system is not a
unique property of Gaussian sources and MSE. It also holds for
symmetrically correlated binary sources with a Hamming distortion.
These sources can be described as follows. Let $X_1, N_1, N_2$ be
mutually independent Ber($1/2$), Ber($p_1$), Ber($p_2$) random
variables respectively. $X_2=X_1\oplus N_1$, $X_3=X_2\oplus N_2$,
where $\oplus$ indicates the Boolean exclusive OR operation. One can
verify that the sum-rate-distortion performance of a C--NC system
matches the JC rate-distortion performance for these sources and
Hamming distortion within a certain distortion region of a nonzero
volume. We omit the proof because it is cumbersome.

\section{Results for the 3-stage NC--C system}\label{sec:NCC}
We can derive the rate-distortion region for an NC--C system by
mimicking the derivations for the C--NC system discussed till this
point. However, due to the operational structural relationship
between C--NC and NC--C systems, it is not necessary to re-derive
the results for the NC--C system at certain operating points, in
particular, for the sum-rate region:

\noindent{\bf Theorem~4}~{\em (``Equivalence'' of C--NC and NC--C
rate-distortion regions)}\\
\noindent{(i)} The rate-distortion region for the one-stage delayed
NC--C system is given by
\begin{eqnarray*}
\mathcal{R}^{NC-C}&=&
\{(\mathbf R, \mathbf D)~|~\exists \ U^2, \widehat X^3, g_1(\cdot),g_2(\cdot,\cdot),s.t.\\
&& R_1 \geq I(X^2;U_1), \\
&& R_2 \geq I(X^3;U_2|U_1), \\
&& R_3 \geq I(X^3;\widehat X_3|U^2), \\
&& D_j \geq E[d_j(X_j,\widehat{X }^j)], \ \ j=1,2,3, \\
&& \widehat X_1 = g_1(U_1),\widehat X_2 = g_2(U_1,U_2),\\
&& U_1 - X^2 - X_3\}.
\end{eqnarray*}
with the following cardinality bounds
\begin{eqnarray*}
|\mathcal U_1| & \leq & |\mathcal X_1| + 6, \\
|\mathcal U_2| & \leq & |\mathcal X_1|^2 |\mathcal X_2|^2|\mathcal
X_3| + 6|\mathcal X_1| |\mathcal X_2||\mathcal X_2| +4.
\end{eqnarray*}

\noindent{(ii)} For an arbitrary distortion tuple $\mathbf{D}$, the
rate regions $\mathcal{R}^{NC-C}$ and $\mathcal{R}^{C-NC}$ are
related in the following manner:
\begin{eqnarray*}
(R_1, R_2, R_3, \mathbf D) \in \mathcal{R}^{C-NC} &\Rightarrow&
(R_1+R_2, R_3,0,\mathbf D) \in \mathcal{R}^{NC-C},\\
(R_1, R_2, R_3,\mathbf D) \in \mathcal{R}^{NC-C} &\Rightarrow& (0,
R_1, R_2+R_3,\mathbf D) \in \mathcal{R}^{C-NC}.
\end{eqnarray*}

\noindent{(iii)} For an arbitrary distortion tuple $\mathbf{D}$, the
minimum sum-rates of one-stage delayed C--NC and NC--C systems are
equal:
\[
R^{C-NC}_{sum}({\bf D})=R^{NC-C}_{sum}({\bf D}).
\]

The proof of part (i) is similar to that of Theorem~3. Part (ii) can
be proved by either using the definitions of $\mathcal{R}^{C-NC}$
and $\mathcal{R}^{NC-C}$ or more directly from the system structure
(see Figs~\ref{fig:otherarch}(a) and (b)) as follows. Given any
C--NC system with rate tuple $(R_1, R_2, R_3)$, we can construct an
NC--C system as follows: (1) combine the first two C--NC encoders to
get the first NC--C encoder, (2) use the third C--NC encoder as the
second NC--C encoder, and (3) use a null encoder with constant zero
output as the third NC--C encoder. Then we have an NC--C system with
rate tuple $(R_1+R_2, R_3, 0)$ and the same distortion tuple.
Similarly, given any NC--C system, we can use a null encoder as the
first C--NC encoder and combine the last two NC--C encoders to get a
C--NC system. Part (iii) follows from part (ii).


The (sum-rate) JC-optimality property of a C--NC system with
one-stage decoding frame-delay given by Corollary~3.2 automatically
holds for an NC--C system with one-stage encoding frame-delay. This
relationship allows one to focus on the performance of only C--NC
systems instead of both C--NC and NC--C systems without loss of
generality. This structural principle holds for the general
multi-frame problem with multi-stage frame-delay, as discussed in
Section~\ref{sec:NCNC}.


\section{General C--NC results}\label{sec:generalresults}
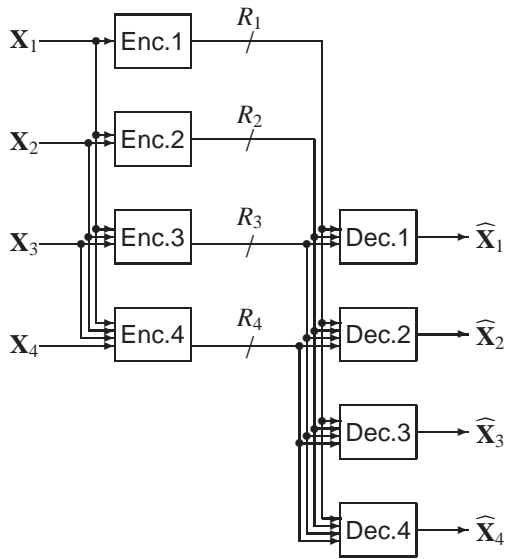
\begin{figure}[!htb]
\centering
\begin{picture}(9,7) 
\put(0.4,2.5){
    \put(2,3.8){\framebox(1.,0.7){\textsf{Enc.1}}}
    \put(2,2.5){\framebox(1.,0.7){\textsf{Enc.2}}}
    \put(2,1.2){\framebox(1.,0.7){\textsf{Enc.3}}}
    \put(2,-0.1){\framebox(1.,0.7){\textsf{Enc.4}}}
    \put(5,1.2){\framebox(1.,0.7){\textsf{Dec.1}}}
    \put(5,-0.1){\framebox(1.,0.7){\textsf{Dec.2}}}
    \put(5,-1.4){\framebox(1.,0.7){\textsf{Dec.3}}}
    \put(5,-2.7){\framebox(1.,0.7){\textsf{Dec.4}}}
    \put(1.0,4.15){\vector(1,0){1}}
    \put(3,4.15){\line(1,0){1.75}}
    \put(1.75,2.9){\vector(1,0){0.25}}
    \put(1.0,2.8){\vector(1,0){1}}
    \put(3,2.85){\line(1,0){1.65}}
    \put(1.0,1.45){\vector(1,0){1}}
    \put(1.75,1.65){\vector(1,0){0.25}}
    \put(1.65,1.55){\vector(1,0){0.35}}
    \put(3,1.45){\vector(1,0){2}}
    \put(4.75,1.65){\vector(1,0){0.25}}
    \put(4.65,1.55){\vector(1,0){0.35}}
    \put(6,1.55){\vector(1,0){0.7}}
    \put(1.55,0.2){\vector(1,0){0.45}}
    \put(1.65,0.3){\vector(1,0){0.35}}
    \put(1.75,0.4){\vector(1,0){0.25}}
    \put(1.0,0.1){\vector(1,0){1}}
    \put(3,0.1){\vector(1,0){2}}
    \put(4.55,0.2){\vector(1,0){0.45}}
    \put(4.65,0.3){\vector(1,0){0.35}}
    \put(4.75,0.4){\vector(1,0){0.25}}
    \put(6,0.25){\vector(1,0){0.7}}
    \put(4.55,-1.1){\vector(1,0){0.45}}
    \put(4.65,-1){\vector(1,0){0.35}}
    \put(4.75,-0.9){\vector(1,0){0.25}}
    \put(4.45,-1.2){\vector(1,0){0.55}}
    \put(6,-1.05){\vector(1,0){0.7}}
    \put(4.55,-2.4){\vector(1,0){0.45}}
    \put(4.65,-2.3){\vector(1,0){0.35}}
    \put(4.75,-2.2){\vector(1,0){0.25}}
    \put(4.45,-2.5){\vector(1,0){0.55}}
    \put(6,-2.35){\vector(1,0){0.7}}
    \put(1.75,4.15){\line(0,-1){3.75}}
    \put(1.65,2.8){\line(0,-1){2.5}}
    \put(1.55,1.45){\line(0,-1){1.25}}
    \put(4.75,4.15){\line(0,-1){6.35}}
    \put(4.65,2.85){\line(0,-1){5.15}}
    \put(4.55,1.45){\line(0,-1){3.85}}
    \put(4.45,0.1){\line(0,-1){2.6}}
    \put(1.75,4.15){\circle*{.1}}
    \put(1.75,2.9){\circle*{.1}}
    \put(1.65,2.8){\circle*{.1}}
    \put(1.75,1.65){\circle*{.1}}
    \put(4.75,1.65){\circle*{.1}}
    \put(1.65,1.55){\circle*{.1}}
    \put(1.55,1.45){\circle*{.1}}
    \put(4.65,1.55){\circle*{.1}}
    \put(4.55,1.45){\circle*{.1}}
    \put(4.75,0.4){\circle*{.1}}
    \put(4.65,0.3){\circle*{.1}}
    \put(4.55,0.2){\circle*{.1}}
    \put(4.45,0.1){\circle*{.1}}
    \put(4.75,-0.9){\circle*{.1}}
    \put(4.65,-1){\circle*{.1}}
    \put(4.55,-1.1){\circle*{.1}}
    \put(4.45,-1.2){\circle*{.1}}
    \put(0.8,4.15){\makebox(0,0){${\bf X}_1$}}
    \put(0.8,2.8){\makebox(0,0){${\bf X}_2$}}
    \put(0.8,1.45){\makebox(0,0){${\bf X}_3$}}
    \put(0.8,0.1){\makebox(0,0){${\bf X}_4$}}
    \put(7,1.55){\makebox(0,0){$\widehat{\bf X}_1$}}
    \put(7,0.25){\makebox(0,0){$\widehat{\bf X}_2$}}
    \put(7,-1.05){\makebox(0,0){$\widehat{\bf X}_3$}}
    \put(7,-2.35){\makebox(0,0){$\widehat{\bf X}_4$}}
    \put(3.8,4.45){\makebox(0,0){$R_1$}}
    \put(3.8,3.15){\makebox(0,0){$R_2$}}
    \put(3.8,1.8){\makebox(0,0){$R_3$}}
    \put(3.8,0.45){\makebox(0,0){$R_4$}}
    \put(3.8,4.14){\makebox(0,0){/}}
    \put(3.8,2.84){\makebox(0,0){/}}
    \put(3.8,1.43){\makebox(0,0){/}}
    \put(3.8,0.08){\makebox(0,0){/}}
    }
\end{picture}
\caption{\label{fig:generalCNC} \small \sl A $4$-stage C--NC system
with a $2$-stage decoding delay.}
\end{figure}

The $T$-stage C--NC system with a $k$-stage decoding delay is a
natural generalization of the 3-stage C--NC system with one-stage
decoding delay. For $j=1,\ldots,T$, encoder-$j$ observes the current
and all the past sources $\mathbf X^j$ and encodes them at rate
$R_j$. Decoder-$j$ observes all the messages sent by encoders one
through $(\min\{j+k,T\})$ and reconstructs $\widehat{\bf {X}}_j$. As
an example, we present the diagram of a C--NC system with $T=4$
frames and $k=2$-stage decoding delay in Fig.~\ref{fig:generalCNC}.
Similar to Theorem~3 and Corollary~3.1, we have the following
results.

{\bf Theorem~5}~{\em (General C--NC rate region)} The rate region
for the $T$-stage C--NC system with a $k$-stage decoding delay is
given by:
\begin{eqnarray*}
\mathcal{R}_k^{C-NC} &=&
\{(\mathbf{R},\mathbf D)~|~\exists \ U^{T-1}, \widehat X^T, g^{T-k-1}(\cdot),s.t. \\
&& R_j \geq I(X^j;U_j|U^{j-1}),\ j=1,\ldots,(T-1)\\
&& R_T \geq I(X^T;\widehat X_{T-k}^T|U^{T-1}), \\
&& U_j - (X^j,U^{j-1}) - X_{j+1}^T,\  j=1,\ldots,(T-1),\\
&& \widehat X_j = g_j(U^{j+k}), \  j=1,\ldots,(T-k-1),\\
&& D_j \geq E[d_j(X_j,\widehat{ X }^j)], \ j=1,\ldots,T\}.
\end{eqnarray*}
with the following cardinality bounds
\[\text{for }j=1,\ldots,T-1,\ \ \ \
|\mathcal U_j| \leq \prod_{k=1}^j |\mathcal
X_k|\prod_{k=1}^{j-1}|\mathcal U_k|+2T.
\]

The cardinality bounds for the alphabets of the auxiliary random
variables stated in Theorem~5 are obtained by a loose counting of
constraints (see Appendix~\ref{appsec:cardinality}). These bounds
 can be improved by some constants by a more careful counting of
constraints. The first term $\prod_{k=1}^j |\mathcal
X_k|\prod_{k=1}^{j-1}|\mathcal U_k|$ comes from the Markov chain
constraints. The second term $2T$ comes from $T$ rate constraints
and $T$ distortion constraints.

{\bf Corollary~5.1}~{\em (General C--NC sum-rate region)} The
sum-rate region of the general C--NC system is given by $\mathcal
R_{k, sum}^{C-NC}({\bf D})=[ R_{k, sum}^{C-NC}({\bf D}),\infty )$,
where $R_{k, sum}^{C-NC}({\bf D})$ is the minimum value of $I(X^T;
\widehat X^T)$ subject to distortion constraints
$E[d_j(X_j,\widehat{ X }^j)] \leq D_j, j=1,\ldots,T$ and Markov
chain constraints
\begin{equation}\label{eqn:CNCMarkovchain}
 \widehat X_j -
(X^{j+k},\widehat X^{j-1}) - X_{j+k+1}^T, \  \ j=1,\ldots,T-k-1.
\end{equation}

For general C--NC systems with increasing system frame-delays, the
expressions of the minimum sum-rates contain the same objective
function $I(X^T;\widehat{X}^T)$ and distortion constraints
$E[d_j(X_j,\widehat{X}^j)] \leq D_j, j = 1,\ldots,T$, but with a
decreasing number of Markov chain constraints. In the limit of
maximum possible system frame-delay, equal to $(T-1)$, which is the
same as in a JC system, we get the JC rate-distortion function with
purely distortion (no Markov chain) constraints. When $T=2$, a
one-stage delayed C--NC system is trivial in terms of the
sum-rate-distortion function because it reduces to that of a 2-stage
JC system. Note that this reduction holds for \emph{arbitrary source
distributions and arbitrary distortion criteria}. So nontrivial
C--NC systems must have at least $T=3$ frames. This is the
motivation for choosing $T=3$ to start the discussion of delayed
sequential coding systems in Section~\ref{sec:structures}. However,
this type of reduction should be distinguished from the nontrivial
reduction result of Corollary~3.2 which only holds for certain
source distributions and distortion criteria.

Using the notation of directed
information~\cite{Massey,Kramerdirect}
\[ I(A^N \rightarrow B^N) := \sum_{n=1}^N I(A^n;B_n|B^{n-1}),
 \]
and its generalization to $k$-directed information~\cite{Prad04}
\begin{eqnarray*}
I_k(A^N \rightarrow B^N) &:=& I(A^N;B^N) - \sum_{n=k+1}^N
I(B^{n-k};A_n|A^{n-1})\\
&=& I(A^N;B^N) - I(0^k B^{N-k} \rightarrow A^N),
\end{eqnarray*}
where $0^k B^{N-k}$ is the $N$-length sequence $(0,\ldots,0,B_1,
\ldots, B_{N-k})$, we can write the objective function of the
minimization problem in Corollary~5.1 as follows
\begin{equation}\label{eqn:kdirect}
 I(X^T; \widehat X^T) =
I_{k+1}(X^T \rightarrow \widehat X^T) + I(0^{k+1}\widehat X^{T-k-1}
\rightarrow X^T).
\end{equation}
The Markov chain constraints (\ref{eqn:CNCMarkovchain}) are
equivalent to the condition $I(0^{k+1}\widehat X^{T-k-1} \rightarrow
X^T) = 0$. So the sum-rate can be reformulated as the minimum of the
first term of (\ref{eqn:kdirect}) subject to the second term $=0$
and the distortion constraints.

As the generalization of Corollary~3.2, we have the following result
for $k$-th order Gauss-Markov sources where $X_1,\ldots,X_T$ form a
$k$-th order Markov chain.

{\bf Corollary~5.2}~{\em (JC optimality of k-stage delayed C--NC
systems for k-th order Gauss-Markov sources and MSE)}
\[R_{k,sum}^{C-NCGM}({\bf D})=R^{JCGM}({\bf D})\]
for the distortion region $\mathcal D^{JC}$.

\begin{proof}
The proof is similar to that of Corollary~3.2. The JC
rate-distortion function is achieved by the test channel
(\ref{eqn:JCtestchannel}) in the distortion region $\mathcal
D^{JC}$. We will verify that the Markov chain $\widehat X_j -
(X^{j+k},\widehat X^{j-1}) - X_{j+k+1}^T$ holds for
$j=1,\ldots,(T-k-1)$.

By the $k$-th order Markov property of the sources, we have $X^j -
X_{j+1}^{j+k} - X_{j+k+1}$. The MMSE estimate of $X_{j+k+1}$ based
on $X^{j+k}$ is given by
\begin{equation}\label{eqn:generalGaussian}
X_{j+k+1} = \sum _{m=1}^k a_{m} X_{j+m} + N\end{equation} where $N$
is a Gaussian random variable which is independent of $X^{j+k}$, and
$\{a_m\}$ are the coefficients of the MMSE estimate. By arguments
which are similar to those used to show the independence of random
variables in the proof of Corollary~3.2, it can be shown that $N$ is
independent of $\widehat X^j$. Therefore the best estimate of
$X_{j+k+1}$ based on $\{X_{j+k},\widehat X^j\}$ is still formula
(\ref{eqn:generalGaussian}). It follows that the Markov chain
$(X^j,\widehat X^j)-X_{j+1}^{j+k}-X_{j+k+1}$ holds which in turn
implies that $\widehat X_j - (X^{j+k},\widehat X^{j-1}) -
X_{j+k+1}^T$ holds and completes the proof.
\end{proof}

This corollary shows that for the $k$-th order Gauss-Markov sources,
the JC sum-rate-MSE performance is achieved by the $k$-stage delayed
C--NC system. Let $\mathcal{D}_d$ denote the distortion region for
which the $d$-stage delayed C--NC sum-rate matches the JC rate for
$k$-th order Gauss-Markov sources and MSE. This region keeps
expanding with delay,
\[\mathcal{D}_k \subseteq \mathcal{D}_{k+1} \subseteq \ldots \subseteq \mathcal{D}_{T-1}=\{\mathbf{R}^+\}^T.\]
The last equality is because the JC system itself has $(T-1)$-stage
delay.

\section{General NC--NC results}\label{sec:NCNC}
\begin{figure}[!htb]
\centering
\begin{picture}(9,7) 
\put(0.4,2.5){
    \put(2,2.5){\framebox(1.,0.7){\textsf{Enc.1}}}
    \put(2,1.2){\framebox(1.,0.7){\textsf{Enc.2}}}
    \put(2,-0.1){\framebox(1.,0.7){\textsf{Enc.3}}}
    \put(2,-1.4){\framebox(1.,0.7){\textsf{Enc.4}}}
    \put(5,1.2){\framebox(1.,0.7){\textsf{Dec.1}}}
    \put(5,-0.1){\framebox(1.,0.7){\textsf{Dec.2}}}
    \put(5,-1.4){\framebox(1.,0.7){\textsf{Dec.3}}}
    \put(5,-2.7){\framebox(1.,0.7){\textsf{Dec.4}}}
    \put(1.0,4.15){\line(1,0){0.75}}
    \put(1.75,2.9){\vector(1,0){0.25}}
    \put(1.0,2.8){\vector(1,0){1}}
    \put(3,2.85){\line(1,0){1.75}}
    \put(1.0,1.45){\vector(1,0){1}}
    \put(1.75,1.65){\vector(1,0){0.25}}
    \put(1.65,1.55){\vector(1,0){0.35}}
    \put(3,1.5){\vector(1,0){2}}
    \put(4.75,1.6){\vector(1,0){0.25}}
    \put(6,1.55){\vector(1,0){0.7}}
    \put(1.55,0.2){\vector(1,0){0.45}}
    \put(1.65,0.3){\vector(1,0){0.35}}
    \put(1.75,0.4){\vector(1,0){0.25}}
    \put(1.0,0.1){\vector(1,0){1}}
    \put(3,0.15){\vector(1,0){2}}
    \put(4.65,0.25){\vector(1,0){0.35}}
    \put(4.75,0.35){\vector(1,0){0.25}}
    \put(6,0.25){\vector(1,0){0.7}}
    \put(1.55,-1.1){\vector(1,0){0.45}}
    \put(1.65,-1){\vector(1,0){0.35}}
    \put(1.75,-0.9){\vector(1,0){0.25}}
    \put(1.45,-1.2){\vector(1,0){0.55}}
    \put(4.55,-1.1){\vector(1,0){0.45}}
    \put(4.65,-1){\vector(1,0){0.35}}
    \put(4.75,-0.9){\vector(1,0){0.25}}
    \put(3,-1.2){\vector(1,0){2}}
    \put(6,-1.05){\vector(1,0){0.7}}
    \put(4.55,-2.4){\vector(1,0){0.45}}
    \put(4.65,-2.3){\vector(1,0){0.35}}
    \put(4.75,-2.2){\vector(1,0){0.25}}
    \put(4.45,-2.5){\vector(1,0){0.55}}
    \put(6,-2.35){\vector(1,0){0.7}}
    \put(1.75,4.15){\line(0,-1){5.05}}
    \put(4.75,2.85){\line(0,-1){5.05}}
    \put(1.65,2.8){\line(0,-1){3.8}}
    \put(1.55,1.45){\line(0,-1){2.55}}
    \put(1.45,0.1){\line(0,-1){1.3}}
    \put(4.65,1.5){\line(0,-1){3.8}}
    \put(4.55,0.15){\line(0,-1){2.55}}
    \put(4.45,-1.2){\line(0,-1){1.3}}
    \put(1.75,2.9){\circle*{.1}}
    \put(1.65,2.8){\circle*{.1}}
    \put(1.75,1.65){\circle*{.1}}
    \put(4.75,1.6){\circle*{.1}}
    \put(1.65,1.55){\circle*{.1}}
    \put(1.55,1.45){\circle*{.1}}
    \put(4.65,1.5){\circle*{.1}}
    \put(1.75,0.4){\circle*{.1}}
    \put(1.65,0.3){\circle*{.1}}
    \put(1.55,0.2){\circle*{.1}}
    \put(1.45,0.1){\circle*{.1}}
    \put(4.75,0.35){\circle*{.1}}
    \put(4.65,0.25){\circle*{.1}}
    \put(4.55,0.15){\circle*{.1}}
    \put(4.75,-0.9){\circle*{.1}}
    \put(4.65,-1){\circle*{.1}}
    \put(4.55,-1.1){\circle*{.1}}
    \put(4.45,-1.2){\circle*{.1}}
    \put(0.8,4.15){\makebox(0,0){${\bf X}_1$}}
    \put(0.8,2.8){\makebox(0,0){${\bf X}_2$}}
    \put(0.8,1.45){\makebox(0,0){${\bf X}_3$}}
    \put(0.8,0.1){\makebox(0,0){${\bf X}_4$}}
    \put(7,1.55){\makebox(0,0){$\widehat{\bf X}_1$}}
    \put(7,0.25){\makebox(0,0){$\widehat{\bf X}_2$}}
    \put(7,-1.05){\makebox(0,0){$\widehat{\bf X}_3$}}
    \put(7,-2.35){\makebox(0,0){$\widehat{\bf X}_4$}}
    \put(3.8,3.15){\makebox(0,0){$R_1$}}
    \put(3.8,1.8){\makebox(0,0){$R_2$}}
    \put(3.8,0.45){\makebox(0,0){$R_3$}}
    \put(3.8,-0.85){\makebox(0,0){$R_4$}}
    \put(3.8,2.84){\makebox(0,0){/}}
    \put(3.8,1.48){\makebox(0,0){/}}
    \put(3.8,0.13){\makebox(0,0){/}}
    \put(3.8,-1.17){\makebox(0,0){/}}
    }
\end{picture}
\caption{\label{fig:NCNC} \small \sl A $4$-stage NC--NC system with
$1$-stage encoding delay and $1$-stage decoding delay. It has the
same sum-rate-distortion performance as the system in
Fig.~\ref{fig:generalCNC}.}
\end{figure}
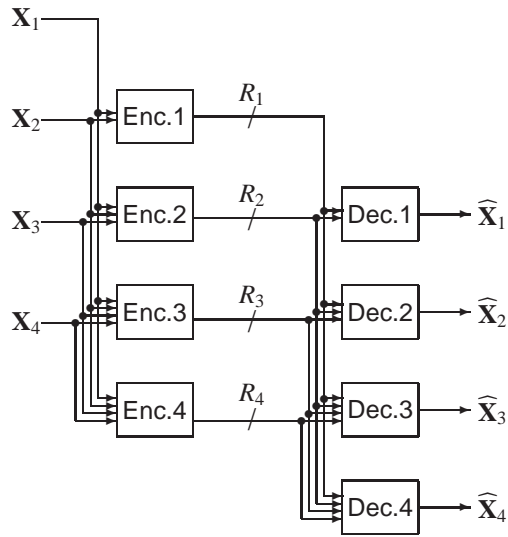

We can consider the general NC--NC systems with $k_1$-stage delay on
the encoder side and $k_2$-stage delay on the decoder side. C--NC
and NC--C systems are special cases when $k_1=0$ and $k_2=0$,
respectively. As an example, in Fig.~\ref{fig:NCNC}, we present the
diagram of an NC--NC system with one-stage encoding delay and
one-stage decoding delay $(T=4, k_1=k_2=1)$. Although NC--NC systems
appear to be structurally more complex, we can relate the
rate-distortion region of NC--NC systems to that of the C--NC
systems using structural arguments as in Section~\ref{sec:NCC}.
Denoting the rate region of the NC--NC systems described above by
$\mathcal R_{k_1,k_2}^{NC-NC}$, we have the following result which
is similar to parts (ii) and (iii) of Theorem~4.

{\bf Theorem~6}~{\em (Relationship between general NC--NC and C--NC
rate regions)} For any distortion tuple $\mathbf D$, \\
\noindent{(i)} $(R_1, \ldots, R_T, \mathbf D) \in
\mathcal{R}^{NC-NC}_{k_1,k_2}
 \Rightarrow
(0,\ldots,0,R_1,\ldots, R_{T-k_1-1},\sum_{j=T-k}^{T}R_j,\mathbf D)
\in \mathcal{R}^{C-NC}_{k_1+k_2}$ \\
\noindent{(ii)} $(R_1, \ldots, R_T,\mathbf D) \in
\mathcal{R}^{C-NC}_{k_1+k_2}
 \Rightarrow  (\sum_{j=1}^{k_1+1} R_j, R_{k_1+2},\ldots,R_T, 0,\ldots,0, \mathbf D) \in
\mathcal{R}^{NC-NC}_{k_1,k_2}.$\\
 where both sequences of zeros
contains $k_1$ zeros.

This result can be proved by noting that the first NC--NC encoder
can be replaced by the combination of the first $k_1$ C--NC
encoders, and the last C--NC encoder can be replaced by the
combination of the last $k_1$ NC--NC encoders, without affecting the
reproduction of frames. As a consequence of this theorem, we have an
exact equivalence between the sum-rates of the NC--NC and the C--NC
systems.

{\bf Corollary~6.1}~{\em (Sum-rate equivalence between NC--NC and
C--NC)} The minimum sum-rates of the $(k_1,k_2)$-stage delayed
NC--NC systems and the $(k_1+k_2)$-stage delayed C--NC systems are
equal:
\[R_{k_1,k_2}^{NC-NC}(\mathbf D) = R_{k_1+k_2}^{C-NC}(\mathbf D).\]

In conclusion, for any two delayed sequential coding systems, when
the {\em sums} of the encoding frame-delay and decoding frame-delay
are equal, they have the same sum-rate-distortion performance. For
example, the NC--NC system in Fig.~\ref{fig:NCNC} has the same
minimum sum-rate as the $2$-stage delayed C--NC system in
Fig.~\ref{fig:generalCNC}. Therefore we can always take the C--NC
system as a representative of all the delayed sequential coding
systems.


\section{Concluding remarks}
In this paper, motivated by video coding applications, we studied
the problem of sequential coding of correlated sources with encoding
and/or decoding frame-delays and characterized the fundamental
tradeoffs between individual frame rates, individual frame
distortions, and encoding/decoding frame-delays in terms of
single-letter information-theoretic quantities. Our characterization
of the rate-distortion region was for multiple sources, general
inter-frame source correlations, and general frame-specific and
coupled single-letter fidelity criteria. The main message of this
study is that even a single frame-delay holds potential for yielding
significant performance improvements in sequential coding problems,
sometimes even matching the joint coding performance.

\section*{Acknowledgment}
The authors would like to thank Prof.~S.~S.~Pradhan, EECS UMich Ann
Arbor and Prof.~K.~Ramchandran, EECS UCBerkeley, for fruitful
discussions and comments.

\appendices
\renewcommand{\theequation}{\thesection.\arabic{equation}}
\setcounter{equation}{0}

\renewcommand{\thedefinition}{\Alph{section}.\arabic{definition}}
\renewcommand{\theremark}{\Alph{section}.\arabic{remark}}
\renewcommand{\thetheorem}{\Alph{section}.\arabic{theorem}}

\renewcommand{\thesubsectiondis}{\Roman{subsection}}
\renewcommand{\thesubsection}{\Alph{section}.\Roman{subsection}}

\section{\label{app:proofcor12}Corollary~1.2 Proof}
We first show that the right hand side of (\ref{eqn:CCGaussian}) is
an upper bound of $R_{sum}^{C-CGM}$ by defining the auxiliary random
variables satisfying all the constraints in (\ref{eqn:CCsumrate})
and evaluating the objective function of (\ref{eqn:CCsumrate}). Then
we show that the right hand side of (\ref{eqn:CCGaussian}) is also
an lower bound of $R_{sum}^{C-CGM}$ using information inequalities.

\emph{Upper bound:} Due to the Markov chains in
(\ref{eqn:CCsumrate}),
\begin{equation}
I(X^3;\widehat X^3)=I(X_1;\widehat X_1)+I(X^2;\widehat X_2|\widehat
X_1)+I(X^3;\widehat X_3|\widehat X^2),\label{eqn:CCGaussian0}
\end{equation}
where on the right hand side, each term corresponds to a stage of
coding. We will sequentially define $\widehat X_1, \widehat X_2,
\widehat X_3$ and evaluate the expression stage by stage to
highlight the structure of the optimal (achievable) coding scheme.

At first, since $D_1\leq \sigma_1^2$, we can find a random variable
$\widehat X_1$ such that: (1) $\widehat X_1 + Z_1 = X_1$, (2)
$\widehat X_1$ and $Z_1$ are independent Gaussian variables with
variances $(\sigma_1^2-D_1)$ and $D_1$ respectively, and (3) the
Markov chain $\widehat X_1- X_1-X_2^3$ holds. The MSE constraint
$E(X_1-\widehat X_1)^2\leq D_1$ is satisfied because $E(Z_1^2)=D_1$.
Note that the distribution of $(X_1,\widehat X_1)$ achieves the
rate-distortion function for the Gaussian source $X_1$ and we have
\begin{equation}\label{eqn:CCGaussian1}
I(X_1;\widehat X_1) = \frac{1}{2}\log
\left(\frac{\sigma_1^2}{D_1}\right).
\end{equation}

Since $(X_1,X_2)$ are jointly Gaussian, we have
\[
X_2 = \rho_1 \frac{\sigma_2}{\sigma_1} X_1 + N_1 = \rho_1
\frac{\sigma_2}{\sigma_1} \widehat X_1 +W_2,
\]
where $N_1$ is a Gaussian variable with variance
$(1-\rho_1^2)\sigma_2^2$ which is independent of $(\widehat
X_1,Z_1)$. $W_2:=\left(\rho_1 \frac{\sigma_2}{\sigma_1} Z_1 +
N_1\right)$ is the innovation from $\widehat X_1$ to $X_2$, whose
variance $\sigma^2_{W_2}$ is given in (\ref{eqn:prederror}). When
$D_2\leq \sigma^2_{W_2}$, we can find a random variable $\widehat
W_2$ such that: (1) $\widehat W_2 + Z_2 = W_2$, (2) $\widehat W_2$
and $Z_2$ are independent Gaussian variables with variances
$(\sigma_{W_2}^2-D_2)$ and $D_2$ respectively, and (3) the Markov
chain $\widehat W_2- (X_2, \widehat X_1)-(X_1,X_3)$ holds. Define
$\widehat X_2:=\left(\rho_1 \frac{\sigma_2}{\sigma_1} \widehat
X_1+\widehat W_2\right)$, which implies $X_2=(\widehat X_2+Z_2)$.
The MSE constraint $E(X_2-\widehat X_2)^2\leq D_2$ is satisfied
because $E(Z_2^2)=D_2$. The Markov chain constraint $\widehat
X_2-(X^2,\widehat X_1)-X_3$ is also satisfied. Note that the
distribution of $(W_2,\widehat W_2)$ achieves the rate-distortion
function for the Gaussian source $W_2$ and we have
\begin{equation}\label{eqn:CCGaussian2}
I(X^2;\widehat X_2|\widehat X_1)=I(X_2;\widehat X_2|\widehat
X_1)=\frac{1}{2}\log \left(\frac{\sigma_{W_2}^2}{D_2}\right),
\end{equation}
where the first step is because $\widehat X_2- (X_2,\widehat
X_1)-X_1$ forms a Markov chain.

Similarly, we can define $\widehat X_3$ such that when $D_3 \leq
\sigma_{W_3}^2$,
\begin{equation}\label{eqn:CCGaussian3}
I(X^3;\widehat X_3|\widehat X^2)=\frac{1}{2}\log
\left(\frac{\sigma_{W_3}^2}{D_3}\right).
\end{equation}

Finally, combining (\ref{eqn:CCGaussian0})-(\ref{eqn:CCGaussian3})
and (\ref{eqn:CCsumrate}), when $\mathbf D\in \mathcal D^{C-C}$, we
have
\[R^{C-CGM}(\mathbf{D}) \leq  \frac{1}{2}\log
\left(\frac{\sigma_1^2}{D_3}\right)+\frac{1}{2}\log
\left(\frac{\sigma_{W_2}^2}{D_2}\right)+\frac{1}{2}\log
\left(\frac{\sigma_{W_3}^2}{D_3}\right).\]

\emph{Lower bound:} For any choice of $\widehat X^3$ satisfying the
constraints in (\ref{eqn:CCsumrate}), we have
\begin{eqnarray}\label{eqn:CCGausslower}
\lefteqn{R^{C-CGM}(\mathbf{D})}\nonumber\\
 &=& \min[ I(X_1;\widehat X_1)+I(X^2;\widehat X_2|\widehat X_1)+I(X^3;\widehat X_3|\widehat X^2)]\nonumber\\
 &\geq& \min [I(X_1;\widehat X_1)+I(X_2;\widehat X_2|\widehat X_1)+I(X_3;\widehat X_3|\widehat X^2)]\nonumber\\
 &=& \min [h(X_1)-h(X_1|\widehat X_1)+h(X_2|\widehat X_1)-h(X_2|\widehat X^2)\nonumber\\
 &&+ h(X_3|\widehat X^2)-h(X_3|\widehat X^3)]\nonumber\\
 & \geq & h(X_1)+\min[h(X_2|\widehat X_1)- h(X_1|\widehat X_1)\nonumber\\
 &&+ h(X_3|\widehat X^2)-h(X_2|\widehat X^2)-h(X_3-\widehat X_3)]\nonumber\\
 & \geq & \frac{1}{2}\log( 2 \pi e \sigma_1^2)-\frac{1}{2}\log (2 \pi e
 D_3) +\min[h(X_2|\widehat X_1)\nonumber\\
 &&-h(X_1|\widehat X_1)]+\min[h(X_3|\widehat X^2)-h(X_2|\widehat
 X^2)],
\end{eqnarray}
where all the minimizations above are subject to all the constraints
in (\ref{eqn:CCsumrate}). By Lemma 5 in \cite{Vis}, since the Markov
chain $\widehat X_1-X_1-X_2$ holds and $h(X_1|\widehat X_1)\leq
\frac{1}{2}\log (2 \pi e D_1)$ we have
\begin{equation}\label{eqn:CCGausslower1}
\min [h(X_2|\widehat X_1)-h(X_1|\widehat X_1)] \geq \frac{1}{2}\log
\left(\frac{\sigma_{W_2}^2}{D_1}\right).\end{equation} Similarly,
since the Markov chain $\widehat X^2-X_2-X_3$ holds and
$h(X_2|\widehat X^2)\leq h(X_2|\widehat X_2)\leq \frac{1}{2} \log (2
\pi e D_2)$, replacing $(X_1, X_2, \widehat X_1)$ by $(X_2, X_3,
\widehat X^2)$ respectively in (\ref{eqn:CCGausslower1}), we have
\begin{equation}\label{eqn:CCGausslower2}
\min [h(X_3|\widehat X^2)-h(X_2|\widehat X^2)] \geq \frac{1}{2}\log
\left(\frac{\sigma_{W_3}^2}{D_2}\right).\end{equation}

Using (\ref{eqn:CCGausslower1}) and (\ref{eqn:CCGausslower2}) in
(\ref{eqn:CCGausslower}), we have
\begin{eqnarray*}
R^{C-CGM}(\mathbf{D})
 & \geq &
\frac{1}{2}\log \left(\frac{ \sigma_1^2}{
 D_3}\right) +\frac{1}{2}\log \left(\frac{\sigma_{W_2}^2}{D_1}\right)+\frac{1}{2}\log
 \left(\frac{\sigma_{W_3}^2}{D_2}\right)\\
& = & \frac{1}{2}\log \left(\frac{ \sigma_1^2}{
 D_1}\right) +\frac{1}{2}\log \left(\frac{\sigma_{W_2}^2}{D_2}\right)+\frac{1}{2}\log
 \left(\frac{\sigma_{W_3}^2}{D_3}\right).
\end{eqnarray*}
In conclusion, the right hand side of the above formula is both an
upper bound and a lower bound of $R^{C-CGM}(\mathbf{D})$, and is
thus equal to $R^{C-CGM}(\mathbf{D})$. \hspace*{\fill}~\QED

\section{\label{app:forwardproofthm3}Theorem~3 forward proof}
For any tuple $(\mathbf R, \mathbf D)$ belonging to the right hand
side of (\ref{eqn:CNCrateregion}), there exist random variables
$U^2, \widehat X^3$ and a function $g_1$ such that all the
constraints in (\ref{eqn:CNCrateregion}) are satisfied. We will
describe the encoders and decoders with parameters $(M^3, R'_2,
R'_3, \epsilon_1)$ in Subsections I to III. In Subsection IV and V,
we choose the values of the parameters and analyze the rates and
distortions to show that (\ref{eqn:admissible1}) and
(\ref{eqn:admissible2}) hold for every $\epsilon > 0$ and
sufficiently large $n$.

\subsection{Generation of codebooks}
\begin{enumerate}
  \item Randomly generate a codebook $\mathcal C_1$ consisting of $M_1$
sequences (codewords) of length $n$ drawn iid $\sim \prod_{i=1}^n
p_{U_1}(u_1(i))$. Index the codewords by $s_1 \in
\{1,2,\ldots,M_1\}$. Denote the $s_1$-th codeword by $\mathbf
U_1(s_1)$.
  \item Randomly generate a codebook $\mathcal C_2$, independently of $\mathcal C_1$, consisting of $2^{nR_2'}$
sequences (codewords) of length $n$ drawn iid $\sim \prod_{i=1}^n
p_{U_2}(u_2(i))$. Index the codewords by $s_2' \in
\{1,2,\ldots,2^{nR_2'}\}$. Denote the $s_2'$-th codeword by $\mathbf
U_2(s_2')$. Then randomly assign the indices of the codewords to one
of $M_2$ bins according to a uniform distribution on
$\{1,2,\ldots,M_2\}$, where $M_2\leq 2^{n R_2'}$. Let $\mathcal
B_2(s_2)$ denote the set of indices assigned to the $s_2$-th bin.
  \item Randomly generate a codebook $\mathcal C_3$, independently of $(\mathcal C_1,\mathcal C_2)$, consisting of $2^{nR_3'}$
sequences (codewords) of length $n$ drawn iid $\sim \prod_{i=1}^n
p_{\hat X_2 \hat X_3}(\hat x_2(i),\hat x_3(i))$. Note that each
component of each codeword is a tuple $(\hat x_2(i),\hat x_3(i)) \in
\widehat {\mathcal X}_2 \times \widehat {\mathcal X}_3$. Index the
codewords by $s_3' \in \{1,2,\ldots,2^{nR_3'}\}$. Denote the
$s_3'$-th codeword by $\widehat {\mathbf X_2^3}(s_3')$. Then
randomly assign the indices to one of $M_3$ bins according to a
uniform distribution on $\{1,2,\ldots,M_3\}$, where $M_3\leq 2^{n
R_3'}$. Let $\mathcal B_3(s_3)$ denote the set of indices assigned
to the $s_3$-th bin.
\end{enumerate}
Reveal all the codebooks to all the encoders and the decoders.

\subsection{Encoding}
\begin{enumerate}
  \item Given a source sequence $\mathbf X_1$, encoder-1 looks for a
  codeword $\mathbf U_1(s_1)$ in $\mathcal C_1$ such that $(\mathbf X_1, \mathbf U_1(s_1))\in
  A_{\epsilon_1}^{*(n)}(p_{X_1 U_1})$ where $\epsilon_1>0$ and $A_{\epsilon_1}^{*(n)}(p_{X_1
  U_1})$ is the $\epsilon_1$-strong typical
  set of length $n$ with respect to the joint distribution $p_{X_1,
  U_1}$\cite{CoverThomas}.  For simplicity, we will
  not indicate either the distribution or the
  length of sequence in the definition of a strong typical set
  if there is no ambiguity. If no such codeword can be found, set $s_1=1$. If more than one
  such codeword exists, pick the one with the smallest index $s_1$. Encoder-1 sends $s_1$
  as the message.

  \item Given sequences $\{\mathbf X_1, \mathbf X_2, \mathbf U_1(s_1)\}$, encoder-2 looks
  for a
  codeword $\mathbf U_2(s_2')$ in $\mathcal C_2$ such that $(\mathbf X_1,\mathbf X_2, \mathbf U_1(s_1), \mathbf U_2(s_2')) \in A_{\epsilon_1}^*$. If no such codeword exists, set $s_2'=1$. If more than one
  such codeword exists, pick the one with the smallest $s_2'$. Encoder-2 sends the bin index $s_2$ such that $s_2' \in \mathcal B_2(s_2)$.

   \item Given sequences $\{\mathbf X_1, \mathbf X_2,\mathbf X_3, \mathbf U_1(s_1),\mathbf U_2(s_2')\}$, encoder-3 looks
  for a
  codeword $\widehat{\mathbf X}_2^3(s_3')$ in codebook $\mathcal C_2$ such that
  $(\mathbf X^3, \mathbf U_1(s_1),\mathbf U_2(s_2'),\widehat{\mathbf X}_2^3(s_3'))\in
  A_{\epsilon_1}^*$. If no such codeword exists, set $s_3'=1$. If more than one
  such codeword exists, pick the one with the smallest $s_3'$. Encoder-3 sends the bin index $s_3$ such that $s_3' \in \mathcal B_3(s_3)$.
\end{enumerate}

\subsection{Decoding}
\begin{enumerate}
  \item Given the received indices $s^2$, decoder-1 looks for a sequence $\mathbf U_2(\hat s_2')$ such that $\hat s_2' \in
  \mathcal B_2(s_2)$ and $(\mathbf U_1(s_1),\mathbf U_2(\hat s_2'))\in
  A_{\epsilon_1}^*$. If more than one such sequence exists,
  pick the one with the smallest $\hat s_2'$. Generate the reproduction sequence $\widehat{\mathbf
  X}_1$ by
  \[\widehat X_1(i) = g_1(U_1(s_1,i),U_2(\hat s_2,i)),\ \ i=1,\ldots,n,\]
  where $\widehat X_1(i), U_1(s_1,i)$, and $U_2(\hat s_2,i)$ are the $i$-th
  components of the sequences $\widehat{\mathbf X}_1, \mathbf U_1(s_1)$, and $\mathbf U_2(\hat
  s_2)$ respectively.
  \item Given the received indices $s^3$ and previously decoded index $\hat
  s_2'$, decoder-2 looks for a sequence $\widehat{\mathbf X}_2^3(\hat s_3')$ such that $\hat s_3' \in
  \mathcal B_3(s_3)$ and $(\mathbf U_1(s_1),\mathbf U_2(\hat s_2'),\widehat{\mathbf X}_2^3(\hat s_3'))\in
  A_{\epsilon_1}^*$. If more than one such sequence exists, pick the one with the smallest
   $\hat s_3'$. Separate $\widehat{\mathbf X}_2^3(\hat s_3')$
  (note that each component is a tuple) to get the reproduction sequences $\widehat{\mathbf X}_2(\hat
  s_3')$ and $\widehat{\mathbf X}_3(\hat s_3')$. This decoder is
  conceptually the combination of decoder-2 and 3 in
  Fig.~\ref{fig:otherarch}(a).
\end{enumerate}

\subsection{Analysis of probabilities of error events}
Let us consider the following ``error events'' $\mathcal E_1$
through $\mathcal E_{11}$. If none of them happens, the decoders
successfully reproduce what the encoders intend to send, and the
expected distortions are closed to $E[d_j(X_j,\widehat X^j)]$, which
is not greater than $D_j$. Otherwise, if any event happens, the
decoders may make mistakes on reproduction, and we will bound the
distortions by the worst case distortion $d_{j,\max}$.

\noindent{$\bullet$}~$\mathcal E_1$: \emph{(Frame-1 not typical)}
$\mathbf X_1 \notin A_{\epsilon_1}^*$.

    $Pr(\mathcal E_1)\rightarrow
  0$ as $n \rightarrow \infty$ by the strong law of large numbers.

\noindent{$\bullet$}~$\mathcal E_2$: \emph{(Encoder-1 fails to find
a codeword)} Given any (deterministic) sequence $\mathbf x_1
\in~A_{\epsilon_1}^*$, $\nexists~ s_1$ such that $(\mathbf x_1,
\mathbf U_1(s_1))\in~A_{\epsilon_1}^*$.

  By \cite[Lemma~13.6.2, p.359]{CoverThomas}, for any
  typical sequence $\mathbf x_1$ and each codeword $\mathbf U_1(s_1)$
  which is randomly generated iid according to $p_{U_1}$, we have
  \[2^{-n(I(X_1;U_1)+\epsilon_2)} \leq Pr((\mathbf x_1,\mathbf U_1(s_1))\in A_{\epsilon_1}^*) \leq 2^{-n(I(X_1;U_1)-\epsilon_2)},\]
  where $\epsilon_2$ depends on $\epsilon_1$ and $n$, and $\epsilon_2\rightarrow 0$ as
  $\epsilon_1 \rightarrow 0$ and
  $n \rightarrow \infty$. Therefore we have
  \begin{eqnarray*}
  Pr(\mathcal E_2) &=& (1-Pr((\mathbf x_1,\mathbf U_1(1))\in
  A_{\epsilon_1}^*))^{M_1}\\
  &{\leq} & \exp\left(-M_1
2^{-n(I(X_1;U_1)+\epsilon_2)}\right),
  \end{eqnarray*}
 where the inequality is because $(1-x)^n \leq \exp(-nx)$.
  Let
  \[M_1 := 2^{n(R_1 + \epsilon_1 +\epsilon_2)}.\]
  Since $R_1\geq I(X_1;U_1)$, we have
\begin{equation*}
  Pr(\mathcal E_2) \leq \exp\left( -2^{n\left(R_1 + \epsilon_1- I(X_1;U_1)\right)}
  \right)
   \leq \exp\left( -2^{n \epsilon_1}  \right)
  \end{equation*}
  which goes to zero as $n\rightarrow \infty$.

\noindent{$\bullet$}~$\mathcal E_3$: \emph{(Message-1 not jointly
typical with
  frame-2)}
  Given any sequences $(\mathbf x_1, \mathbf u_1(s_1))\in
  A_{\epsilon_1}^*$, $(\mathbf x_1, \mathbf X_2, \mathbf u_1(s_1)) \notin
  A_{\epsilon_1}^*$.

  Using the
  Markov lemma \cite[Lemma~14.8.1, p.436]{CoverThomas}, since the Markov chain $U_1 - X_1 - X_2$ holds and $\mathbf
  X_2$ is drawn iid $\sim p_{X_2|X_1}$, $Pr((\mathbf x_1, \mathbf X_2, \mathbf u_1(s_1)) \notin
  A_{\epsilon_1}^*) \leq \epsilon_1$ for $n$ sufficiently large. Therefore $Pr(\mathcal E_3)\rightarrow
  0$ as $\epsilon_1 \rightarrow 0$, and $n \rightarrow \infty$.

\noindent{$\bullet$}~$\mathcal E_4$: \emph{(Encoder-2 fails to find
a codeword)}
  Given any sequences $(\mathbf x_1, \mathbf x_2, \mathbf
  u_1(s_1))\in A_{\epsilon_1}^*$, $\nexists~s_2'$ such that $(\mathbf x_1,\mathbf x_2, \mathbf u_1(s_1),$ $\mathbf U_2(s_2'))\in  A_{\epsilon_1}^*$.

  By arguments which are similar to those used in the analysis of $\mathcal E_2$, we have
    \[Pr(\mathcal E_4) \leq \exp\left( -2^{n(R_2' - I(X^2 U_1;U_2) - \epsilon_3)}\right), \]
    where $\epsilon_3 \rightarrow 0$ as $\epsilon_1 \rightarrow 0$ and
  $n \rightarrow \infty$.
  Let
  \[R_2' := I(X^2 U_1;U_2) + \epsilon_1+ \epsilon_3.\]
  We have
  \[ Pr(\mathcal E_4) \leq \exp\left( -2^{n \epsilon_1}  \right), \]
  which goes to zero as $n\rightarrow \infty$.

\noindent{$\bullet$}~$\mathcal E_5$: \emph{(Encoder-2's bin size too
large)}
    Given that $s_2' \in \mathcal B_2(s_2)$, the cardinality of the $s_2$-th bin satisfies
    \[ |\mathcal B_2(s_2)| \geq \frac{2^{n(R_2'+\epsilon_1)}}{M_2}+1.\]

    Because $s_2' \in \mathcal B_2(s_2)$ and the other $\left(2^{nR_2'}-1\right)$ codewords are randomly assigned,
    $(|\mathcal B_2(s_2)|-1)$ follows the binomial distribution with parameters
    $(2^{nR_2'}-1,1/M_1)$. We will use the following Chernoff
    bound\cite[Thm~4.4(3), p.64]{Mitzenmacher}: For a binomial random
    variable $X$ with parameters $(n,p)$, if $a \geq 6np$, then $Pr(X \geq a) \leq
    2^{-a}$. When $n \epsilon_1 > 3$ which guarantees $2^{n \epsilon_1}>
    6$, taking $a:=2^{R_2'+\epsilon_1}/M_2$, we have
    \[ Pr \left(|\mathcal B_2(s_2)|-1\geq \frac{2^{n (R_2'+\epsilon_1)}}{M_2}\right) \leq 2^ {- \frac{2^{n (R_2'+\epsilon_1)}}{M_2}}.\]
    Since $M_2 \leq 2^{n R_2'}$, we have $Pr(\mathcal E_5) \rightarrow
    0$ as $n \rightarrow \infty$.

\noindent{$\bullet$}~$\mathcal E_6$: \emph{(Decoder-1 fails to
identify the correct codeword from the
  bin)} In the bin $\mathcal B_2(s_2)$ whose size is not greater than $\left(2^{n(R_2'+\epsilon_1)}/M_2+1\right)$, Given any sequence $\mathbf
  u_1(s_1)\in A_{\epsilon_1}^*$,
   $\exists~\hat s_2' \neq s_2'$ such that $(\mathbf u_1(s_1), \mathbf U_2(\hat s_2'))\in
   A_{\epsilon_1}^*$.

  By arguments which are similar to those used in the analysis of $\mathcal E_2$, we have
  \[Pr((\mathbf u_1(s_1), \mathbf U_2(\hat s_2'))\in A_{\epsilon_1}^*) \leq 2^{-n
  (I(U_1;U_2)-\epsilon_4)},\]
  where $\epsilon_4 \rightarrow 0$ as $\epsilon_1 \rightarrow 0$ and
  $n \rightarrow \infty$. By the union bound,
    \begin{eqnarray*} Pr(\mathcal E_6) &\leq& (|\mathcal B_2(s_2)|-1) 2^{-n
  (I(U_1;U_2)-\epsilon_4)} \\
  &\leq& 2^{n
  \left(R_2'+\epsilon_1 -
  I(U_1;U_2)+\epsilon_4\right)}/M_2.\end{eqnarray*}
Recall that
\[R_2' = I(X^2 U_1;U_2) + \epsilon_1+\epsilon_3.\]
Let
\[M_2 := 2^{n(R_2 + 3\epsilon_1 +\epsilon_3+\epsilon_4)}.\]
Due to the fact that $R_2 \geq I(X^2;U_2|U_1)$, we can simplify the
bound to
\[Pr(\mathcal E_6) \leq 2^{n
  \left(I(X^2;U_2|U_1) - R_2 - \epsilon_1 \right)} \leq 2^{-n \epsilon_1},\]
which goes to zero as $n \rightarrow \infty$.

\noindent{$\bullet$}~$\mathcal E_7$: \emph{(Frame-3 not jointly
typical with previous messages)} Given any sequences $(\mathbf x_1,
\mathbf x_2, \mathbf u_1(s_1), \mathbf u_2(s_2'))\in
A_{\epsilon_1}^*$, $(\mathbf x_1, \mathbf x_2, \mathbf u_1(s_1),
\mathbf u_2(s_2'),\mathbf X_3)\notin A_{\epsilon_1}^*$.

By arguments which are similar to those used in the analysis of
$\mathcal E_3$, the Markov chain $U^2 - X^2 - X_3$ implies that
$Pr(\mathcal E_7) \rightarrow 0$ as $\epsilon_1 \rightarrow 0$ and
$n \rightarrow \infty$.

\noindent{$\bullet$}~$\mathcal E_8$: \emph{(Encoder-3 fails to find
a codeword)} Given any sequences $(\mathbf x_1, \mathbf x_2, \mathbf
x_3, \mathbf
  u_1(s_1),\mathbf u_2(s_2'))\in A_{\epsilon_1}^*$, $\nexists~s_3'$
  such that $(\mathbf x_1,\mathbf x_2, \mathbf x_3, \mathbf u_1(s_1), \mathbf u_2(s_2'), \widehat{\mathbf X_2^3}(s_3'))\in
  A_{\epsilon_1}^*$.

  By arguments which are similar to those used in the analysis of $\mathcal E_2$, we
  have
    \[Pr(\mathcal E_8) \leq \exp\left( -2^{n(R_3' - I(X^3 U^2;\widehat X_2^3) - \epsilon_5)}\right), \]
    where $\epsilon_5 \rightarrow 0$ as $\epsilon_1 \rightarrow 0$ and
  $n \rightarrow \infty$.
  Let
  \[R_3' := I(X^3 U^2;\widehat X_2^3) + \epsilon_1+\epsilon_5.\]
  We have
  \[ Pr(\mathcal E_8) \leq \exp\left( -2^{n \epsilon_1} \right), \]
  which goes to zero when $n\rightarrow \infty$.

\noindent{$\bullet$}~$\mathcal E_9$: \emph{(Encoder-3's bin size too
large)} Given that $s_3' \in \mathcal B_3(s_3)$, the cardinality of
the bin satisfies
\[|\mathcal B_3(s_3)| > \frac{2^{n(R_3'+\epsilon_1)}}{M_3}+1.\]
By arguments which are similar to those used in the analysis of
$\mathcal E_5$, we can argue that $Pr(\mathcal E_9) \rightarrow 0$
as $n \rightarrow \infty$.

\noindent{$\bullet$}~$\mathcal E_{10}$: \emph{(Decoder-2 fails to
identify the correct codeword from the bin)} In the bin $\mathcal
B_3(s_3)$ whose size is not greater than
$\left(2^{n(R_3'+\epsilon_1)}/M_3+1\right)$, given any sequences
$(\mathbf u_1(s_1), \mathbf u_2(s_2'))\in A_{\epsilon_1}^*$,
$\exists~\hat s_3' \neq s_3'$ such that $(\mathbf u_1(s_1), \mathbf
u_2(s_2'), \widehat{\mathbf X}_2^3(\hat s_3')) \in
A_{\epsilon_1}^*$.

By arguments which are similar to those used in the analysis of
$\mathcal E_6$, we have
\[ Pr(\mathcal E_{10}) \leq 2^{n
  \left(I(X^3;\hat X_2^3|U^2) +2 \epsilon_1 +\epsilon_5+\epsilon_6\right)}/M_3,\]
where $\epsilon_6 \rightarrow 0$ as $\epsilon_1 \rightarrow 0$ and
  $n \rightarrow \infty$.
Let
 \[M_3 := 2^{n(R_3 + 3\epsilon_1 +\epsilon_5+\epsilon_6)}.\]
Due to the fact that $R_3\geq I(X^3;\widehat X_2^3|U^2)$, we have
\[Pr(\mathcal E_{10}) \leq 2^{n \epsilon_1},\]
which goes to zero as $n \rightarrow \infty$.

\noindent{$\bullet$}~$\mathcal E_{11}$: \emph{(Reproduction of
frame-1 not jointly typical with other sequences)} Given any
sequences $(\mathbf x^3, \hat {\mathbf x}_2^3(s_3'), \mathbf
u_1(s_1), \mathbf u_2(s_2'))\in A_{\epsilon_1}^*$ and a correct
decoding $\hat s_2'=s_2'$, $(\mathbf x^3, \hat{\mathbf x}_2^3(s_3'),
\mathbf u_1(s_1), \mathbf u_2(s_2'),\hat{\mathbf x}_1)\notin
A_{\epsilon_1}^*$.

Although $\hat {\mathbf x}_1$ depends on $(\mathbf u_1(s_1),\mathbf
u_2(\hat s_2'))$ deterministically by the function $g_1$, we can
regard $p_{\hat X_1|U_1,U_2}$ as a degraded probability distribution
and use the Markov lemma and the trivial Markov chain $(X^3,\widehat
X_2^3) - U^2 - g_1(U^2)$ to show that $Pr(\mathcal E_{11})
\rightarrow 0$ as $\epsilon_1 \rightarrow 0$ and $n \rightarrow
\infty$.

\subsection{Analysis of the distortions}

 Consider the union
of all the above events $\mathcal E :=\bigcup_{i=1}^{11}\mathcal
E_{i}$. When the codebooks are randomly generated according to
Subsection~I, since $Pr(\mathcal E_i)$ vanishes for $i=1,\ldots,11$
as $\epsilon_1 \rightarrow 0$, and $n \rightarrow \infty$,
$Pr(\mathcal E)$ also vanishes. Therefore there must exists a
sequence of codebooks $\{(\mathcal C_{1,l},\mathcal C_{2,l},\mathcal
C_{3,l})\}_{l=1}^\infty$ for which $Pr(\mathcal E) \rightarrow 0$
(the randomness comes from the generation of source sequences). We
will focus on these codebooks in the following discussion.

In the case that $\mathcal E$ does not happen, all the sequences are
jointly $\epsilon_1$-strong typical: $(\mathbf X^3, \mathbf
U_1(s_1), \mathbf U_2(s_2'),\widehat{\mathbf X}_1, \widehat{\mathbf
X}_2^3(s_3'))\in A_{\epsilon_1}^*$, and the decoded indices are
correct: $\hat s_2'=s_2', \hat s_3'=s_3'$. Since the expected
distortion is a continuous function of the joint distribution,
strong typicality implies distortion typicality. In other words, we
have
\[ |E[d_j^{(n)}(\mathbf X_j, \widehat{\mathbf X}^j)|\mathcal E^c] - E[d_j(X_j, \widehat X^j)]| < \epsilon_{d_j}, \ j=1,2,3,\]
where $\epsilon_{d_j} \rightarrow 0$ as $\epsilon_1 \rightarrow 0$.
Since $D_j \geq E[ d_j(X_j, \widehat X^j)]$, we have
$E[d_j^{(n)}(\mathbf X_j, \widehat{\mathbf X}^j)|\mathcal E^c]\leq
D_j+ \epsilon_{d_j}, j=1,2,3$.

In the case that $\mathcal E$ does happen, by the definition of
$d_{j,\max}$, we have $E[d_j^{(n)}(\mathbf X_j, \widehat{\mathbf
X}^j)|\mathcal E]\leq d_{j,\max}, j=1,2,3$.

Therefore the expected distortion for the $j$-th frame is,
\begin{eqnarray*}
E[d_j^{(n)}(\mathbf X_j, \widehat{\mathbf X}^j)]&=&
E[d_j^{(n)}(\mathbf X_j, \widehat{\mathbf X}^j)|\mathcal E]
Pr(\mathcal E)\\&& + E[d_j^{(n)}(\mathbf X_j,
\widehat{\mathbf X}^j)|\mathcal E^c] (1 - Pr({\mathcal E}))\\
& \leq & d_{j,\max} Pr(\mathcal E) + E[d_j(\mathbf X_j,
\widehat{\mathbf X}^j)|\mathcal E^c]\\
& \leq & d_{j,\max} Pr(\mathcal E) + D_j+ \epsilon_{d_j}
\end{eqnarray*}

When $\epsilon_1 \rightarrow 0$, and $n \rightarrow \infty$, for
codebooks $\{(\mathcal C_{1,l},\mathcal C_{2,l},\mathcal
C_{3,l})\}_{l=1}^\infty$, $Pr(\mathcal E)$ and all the $\epsilon$
variables vanish. Therefore $\forall \epsilon >0$, by driving the
variables to their limits, we can always find $\epsilon_1 > 0$ for
sufficiently large $n$, such that
\begin{eqnarray*}
\frac{1}{n}\log M_1 - R_1 &=&  \epsilon_1 +\epsilon_2 < \epsilon,\\
\frac{1}{n}\log M_2 - R_2 &=& 3 \epsilon_1 +\epsilon_3+\epsilon_4 <
\epsilon,\\
\frac{1}{n}\log M_3 - R_3 &=& 3 \epsilon_1 +\epsilon_5+\epsilon_6 <
\epsilon,\\
 E[d_j^{(n)}(\mathbf X_j, \widehat{\mathbf X}^j)]- D_j &\leq&
  d_{j,\max} Pr(\mathcal E) + \epsilon_{d_j} < \epsilon.
\end{eqnarray*}
Therefore (\ref{eqn:admissible1}) and (\ref{eqn:admissible2}) hold,
which completes the proof. \hspace*{\fill}~\QED

\section{\label{app:proofthm3}Theorem~3 converse proof}
\subsection{Information equalities}
If a rate-distortion-tuple $(\mathbf R, \mathbf D)= (R_1, \ldots,
R_T , D_1, \ldots, D_T)$ is admissible for the 3-stage C--NC system,
then $\forall \epsilon > 0$, there exists $N(\epsilon)$, such that
$\forall n> N(\epsilon)$ we have blocklength $n$ encoders and
decoders $\{f_1^{(n)}, f_2^{(n)}, f_3^{(n)}, g_1^{(n)}, g_2^{(n)},
g_3^{(n)}\}$ satisfying
\begin{eqnarray*}
E[d_j(\mathbf X_j,\widehat{ \bf{X} }^j)] & \leq & D_j+\epsilon,\\
\frac{1}{n} \log M_j & \leq & R_j +\epsilon,
 \ \ j=1,\ldots,T.
\end{eqnarray*}

Denote the messages sent by the three ($T = 3$) encoders
respectively by $S_1, S_2,$ and $S_3$, and define the auxiliary
random variables by $U_j(i):=(S_j,X_j(i-)), j=1,2$. Due to the
structure of the system we have the following Markov chains
\begin{eqnarray*}
&\mathbf X_2^3 - \mathbf X_1- S_1,&\\
&\mathbf X_3 - \mathbf X^2- S^2- \widehat{\mathbf X}_1,&
\end{eqnarray*}
which are readily verified. For the first coding rate, we have
\begin{eqnarray*}
n(R_1+\epsilon) &\geq& H(S_1)\\
&=&H(S_1)-H(S_1|\mathbf X_1)\\
&=&I(S_1;\mathbf X_1)\\
&=&\sum_{i=1}^n I(X_1(i);S_1|X_1(i-))\\
&\stackrel{(a)}{=}&\sum_{i=1}^n I(X_1(i);S_1,X_1(i-))\\
&=&\sum_{i=1}^n I(X_1(i);U_1(i))
\end{eqnarray*}

Step (a) is because $(X_1(i))_{i=1}^n$ are iid. The Markov chains
$X_2^3(i)-X_1(i)-U_1(i)$ can be verified to hold for each
$i=1,\ldots,n$.

 In the next stage,
\begin{eqnarray*}
n(R_2+\epsilon) &\geq& H(S_2)\\
&\geq& H(S_2|S_1)\\
&=& H(S_2|S_1)-H(S_2|S_1,\mathbf X^2)\\
&=& I(S_2;\mathbf X^2|S_1)\\
&=& \sum_{i=1}^n I(X^2(i);S_2|U_1(i),X_2(i-))\\
&\stackrel{(b)}{=}& \sum_{i=1}^n I(X^2(i);S_2,X_2(i-)|U_1(i))\\
&=& \sum_{i=1}^n I(X^2(i);U_2(i)|U_1(i))
\end{eqnarray*}
Step (b) is because the Markov chain $X^2(i)-U_1(i)-X_2(i-)$ holds
for each $i$. For each $i$, $\widehat X_1(i)$ is a deterministic
function of $S^2$, which is itself a deterministic function of
$U^2(i)$. Therefore there exists a function $g_{1,i}$ such that
$\widehat X_1(i) = g_{1,i}(U^2(i))$ for each $i$. The Markov chain
$X_3(i)-(X^2(i),U_1(i))-U_2(i)$ can also be verified to hold for
each $i=1,\ldots,n$.

In the final stage,
\begin{eqnarray*}
n(R_3+\epsilon) &\geq& H(S_3)\\
&\geq& H(S_3|S^2)\\
&=& H(S_3|S^2)-H(S_3|S^2,\mathbf X^3)\\
&=& I(S_3;\mathbf X^3|S^2)\\
&=& \sum_{i=1}^n I(X^3(i);S_3|U^2(i),X_3(i-))\\
&\stackrel{(c)}{=}& \sum_{i=1}^n I(X^3(i);S_3,X_3(i-)|U^2(i))\\
&\stackrel{(d)}{\geq}& \sum_{i=1}^n I(X^3(i);\widehat
X_2^3(i)|U^2(i))
\end{eqnarray*}
where step (c) is because the Markov chain $X^3(i)-U^2(i)-X_3(i-)$
holds for each $i$. Step (d) is because $\widehat X_2^3(i)$ is a
deterministic function of $\{S_1, S_2, S_3\} \subseteq \{ S_3,
U_1(i), U_2(i)\}$ for each $i=1,\ldots,n$.

Hence we have shown that for any admissible rate-distortion tuple
$(\mathbf R, \mathbf D)$, $\forall \epsilon>0, \exists~N(\epsilon)$
such that for all $n > N(\epsilon)$,
\begin{eqnarray*}
&&R_1+\epsilon \geq \frac{1}{n} \sum_{i=1}^n I(X_1(i);U_1(i)),\\
&&R_2+\epsilon \geq \frac{1}{n} \sum_{i=1}^n I(X^2(i);U_2(i)|U_1(i)),\\
&&R_3+\epsilon \geq \frac{1}{n} \sum_{i=1}^n I(X^3(i);\widehat X_2^3(i)|U^2(i)),\\
&&D_j+\epsilon \geq E[d_j^{(n)}(\mathbf X_j, \widehat
{\mathbf X}^j)], \ \ j=1,\ldots,T,\\
&&\widehat X_1(i) = g_{1,i}(U^2(i)), i=1,\ldots,n
\end{eqnarray*}
and the Markov chains $X_2^3(i)-X_1(i)-U_1(i)$ and
$X_3(i)-(X^2(i),U_1(i))-U_2(i)$ hold for each $i$. Note that the
Markov chains imply that
\begin{eqnarray}
&&\sum_{i=1}^n I(U_1(i);X_2^3(i)|X_1(i))=0,\label{eqn:Timesharingthm3_Markov1}\\
&&\sum_{i=1}^n I(U_2(i);X_3(i)|X^2(i),U_1(i))=0.
\label{eqn:Timesharingthm3_Markov2}
\end{eqnarray}

\subsection{Time-sharing}\label{appsec:timesharing}
We introduce a timesharing random variable $Q$ taking values in
$\{1,\ldots,n\}$ equally likely, which is independent of all the
other random variables. We have
\begin{eqnarray*}
R_1+\epsilon & \geq & \frac{1}{n} \sum_{i=1}^n I(X_1(i);U_1(i))\\
& = & \frac{1}{n} \sum_{i=1}^n I(X_1(i);U_1(i)|Q=i)\\
&=&I(X_1(Q);U_1(Q)|Q)\\ &=& I(X_1(Q);U_1(Q),Q)
\end{eqnarray*}
Similarly, we have
\[
R_2+\epsilon \geq I(X^2(Q);U_2(Q)|U_1(Q),Q),
\]
\[
R_3+\epsilon \geq I(X^3(Q);\widehat X_2^3(Q)|U^2(Q),Q),
\]
\[
D_j+\epsilon \geq E[ d_j (X_j(Q), \widehat X^j(Q))].
\]
Now define $U_1 := (U_1(Q),Q)$, $U_2 :=U_2(Q)$, $X_j := X_j(Q)$,
$\widehat X_j := \widehat X_j(Q)$ for $j=1,\ldots,T$. Also define
deterministic functions $g_1$ as follows,
\[
g_1(U^2)=g_1(U^2(Q),Q):=g_{1,Q}(U^2(Q)) = \widehat X_1(Q)=\widehat X_1,\\
\]
which are consistent with the definitions of $\{U^2, \widehat
X_1\}$. Then we have the inequalities
\begin{eqnarray}
&&R_1+\epsilon \geq I(X_1;U_1),\label{eqn:Timesharingthm3_1}\\
&&R_2+\epsilon \geq I(X^2;U_2|U_1),\label{eqn:Timesharingthm3_2}\\
&&R_3+\epsilon \geq I(X^3;\widehat X_2^3|U^2),\label{eqn:Timesharingthm3_3}\\
&&D_j+\epsilon \geq E[d_j(X_j, \widehat X^j)], \ \
j=1,\ldots,T.\label{eqn:Timesharingthm3_4}
\end{eqnarray}
Concerning the Markov chains, note that
\begin{eqnarray*}
I(U_1;X_2^3|X_1)& = & I(U_1(Q),Q;X_2^3(Q)|X_1(Q))\\
&=& I(Q; X_2^3(Q)|X_1(Q)) \\
&&+ I(U_1(Q);X_2^3(Q)|X_1(Q),Q).
\end{eqnarray*}
The first term is zero because $Q$ is designed to be independent of
all other random variables. The second term is zero because of
Equation~(\ref{eqn:Timesharingthm3_Markov1}). Hence the Markov chain
$U_1-X_1-X_2^3$ holds. Furthermore, we have
\[
I(U_2;X_3|X^2,U_1)= I(U_2(Q);X_3(Q)|X^2(Q),U_1(Q),Q)=0,
\]
because of Equation~(\ref{eqn:Timesharingthm3_Markov2}). Hence the
Markov chain $U_2-(X^2,U_1)-X_3$ holds.

\subsection{Cardinality bounds on the alphabet of auxiliary random variables}
\label{appsec:cardinality}
 Till now we have shown that for any
admissible rate-distortion tuple $(\mathbf R, \mathbf D)$, $\forall
\epsilon >0$, for sufficiently large $n$, inequalities
(\ref{eqn:Timesharingthm3_1})-(\ref{eqn:Timesharingthm3_4}) and the
Markov chains $U_1-X_1-X_2^3$ and $U_2-(X^2,U_1)-X_3$ hold. The
definition of $U_j(i)=(S_j, X_j(1),\ldots,X_j(i-1))$ guarantees that
$U_j(i)$ has a finite alphabet, although its cardinality grows with
$n$. Therefore $U_1=(U_1(Q),Q)$, $U_2=U_2(Q)$ also have the finite
alphabets $\mathcal U_1$ and $\mathcal U_2$ whose cardinalities grow
with $n$. In this section, we will use the Carath\'eodory theorem to
find new random variables $U_1^*$ and $U_2^{**}$ with smaller
alphabets whose sizes are independent of $n$, such that inequalities
(\ref{eqn:Timesharingthm3_1})-(\ref{eqn:Timesharingthm3_4}) and the
Markov chains still hold even if $\{U_1, U_2\}$ are replaced by
$\{U_1^*, U_2^{**}\}$.

Observe that we can define functionals $\{f_{x_1}\}_{x_1 \in
\mathcal X_1}, f_{R_j}, f_{d_j},j=1,2,3$ as follows. Note that they
depend on $u_1$, conditional distributions conditioned on $U_1$ and
the function $g_1$.

\begin{eqnarray}\label{eqn:CardXthm3_1}
p_{X_1}(x_1) &=& \sum_{u_1 \in \mathcal U_1} p_{U_1}(u_1) p_{X_1|U_1}(x_1|u_1)\nonumber \\
&=:& \sum_{u_1 \in \mathcal U_1} p_{U_1}(u_1)
f_{x_1}(u_1,p_{X_1|U_1}), \forall x_1 \in \mathcal X_1,
\end{eqnarray}
\begin{eqnarray}\label{eqn:CardRthm3_1}
I(X_1;U_1) & = & H(X_1) - \sum_{u_1 \in \mathcal U_1} p_{U_1}(u_1) H(X_1|U_1=u_1) \nonumber \\
 & =:& \sum_{u_1 \in \mathcal U_1} p_{U_1}(u_1)
 f_{R_1}(u_1,p_{X_1|U_1}),
\end{eqnarray}
\begin{eqnarray}\label{eqn:CardRthm3_2}
I(X^2;U_2|U_1) & = & \sum_{u_1 \in \mathcal U_1} p_{U_1}(u_1) I(X^2;U_2|U_1=u_1) \nonumber \\
 & =: & \sum_{u_1 \in \mathcal U_1} p_{U_1}(u_1)
 f_{R_2}(u_1,p_{X^2U_2|U_1}),
\end{eqnarray}
\begin{eqnarray}\label{eqn:CardRthm3_3}
I(X^3;\widehat X_2^3|U_1,U_2)  &=&  \sum_{u_1 \in \mathcal U_1} p_{U_1}(u_1) I(X^3;\widehat X_2^3|U_1=u_1,U_2) \nonumber \\
 & =:& \sum_{u_1 \in \mathcal U_1} p_{U_1}(u_1) f_{R_3}(u_1,p_{X^3U_2\hat
 X_3|U_1}),
\end{eqnarray}
\begin{eqnarray}\label{eqn:CardDthm3_1}
E[d_1(X_1,\widehat X_1)] & = & \sum_{u_1 \in \mathcal U_1} p_{U_1}(u_1) E[d_1(X_1,g_1(u_1,U_2))|U_1=u_1] \nonumber \\
 & =: & \sum_{u_1 \in \mathcal U_1} p_{U_1}(u_1)
 f_{d_1}(u_1,p_{X_1 U_2|U_1},g_1),
\end{eqnarray}
\begin{eqnarray}\label{eqn:CardDthm3_2}
\lefteqn{E[d_2(X_2,\widehat X^2)]}\nonumber \\
& = & \sum_{u_1 \in \mathcal U_1} p_{U_1}(u_1) E[d_2(X_2,g_1(u_1,U_2),\widehat X_2|U_1=u_1] \nonumber \\
 & =: & \sum_{u_1 \in \mathcal U_1} p_{U_1}(u_1)
 f_{d_2}(u_1,p_{X_2U_2\widehat X_2|U_1},g_1),
\end{eqnarray}
\begin{eqnarray}\label{eqn:CardDthm3_3}
\lefteqn{E[d_3(X_3,\widehat X^3)]}\nonumber \\
& = & \sum_{u_1 \in \mathcal U_1} p_{U_1}(u_1) E[d_3(X_3,g_1(u_1,U_2),\widehat X_2^3)|U_1=u_1] \nonumber \\
 & =: & \sum_{u_1 \in \mathcal U_1} p_{U_1}(u_1) f_{d_3}(u_1,p_{X_3U_2\hat
 X_2^3|U_1},g_1).
\end{eqnarray}

We try to find a new random variable $U_1^*$ to replace $U_1$ such
that all the quantities in the above equations need to be preserved.
Because the Markov chains $U_1 - X_1 - X_2^3$ and $U_2 - (X^2,U_1) -
X_3$ hold, we can write the joint distribution as follows,
\[ p_{X^3U^2\hat X_2^3} = p_{U_1} p_{X_1|U_1} p_{X_2 X_3|X_1} p_{U_2 |X^2 U_1} p_{\hat X_2^3|X^3 U^2}.\]
Fixing $p_{X_1|U_1}, p_{X_2 X_3|X_1}, p_{U_2 |X^2 U_1}, p_{\hat
X_2^3|X^3 U^2}$ and $g_1$, the functionals $\{f_{x_1}\}_{x_1 \in
\mathcal X_1}, f_{R_j}, f_{d_j},j=1,2,3$ become functions depending
solely on $u_1$.

Since $p_{X_1}(x_1)$ is a probability mass function which always
adds up to $1$, we only care about $(|\mathcal X_1|-1)$ out of
$|\mathcal X_1|$ equations~(\ref{eqn:CardXthm3_1}). Suppose
$\{x_{1,1},\ldots,x_{1,|\mathcal X_1|-1}\}\subset \mathcal X_1$ are
$(|\mathcal X_1|-1)$ different elements of interest. Consider a set
of $k=|\mathcal X_1|+5$ dimensional vectors consisting of $|\mathcal
U_1|$ elements
\begin{eqnarray*}
\mathcal A = \{(f_{x_{1,1}}(u_1),\ldots,f_{x_{1,|\mathcal
X_1|-1}}(u_1),f_{R_1}(u_1),f_{R_2}(u_1),\\f_{R_3}(u_1),
f_{d_1}(u_1),f_{d_2}(u_1),f_{d_3}(u_1))\}_{u_1 \in \mathcal U_1}.
\end{eqnarray*}
According to the above equations, the vector
\begin{eqnarray*}
\mathbf a&  =& (p_{X_1}(x_{1,1}),\ldots,p_{X_1}(x_{1,|\mathcal
X_1|-1}),I(X_1;U_1),I(X^2;U_2|U_1),\\
&& \!\!\!\!\!\! I(X^3;\widehat X_3|U^2), E[d_1(X_1,\widehat
X_1)],E[d_2(X_2,\widehat X^2)],E[d_3(X_3,\widehat X^3)])
\end{eqnarray*}
is in the convex hull of set $\mathcal A$. By the Carath\'eodory
theorem \cite{Eggleston}, there exist $(k+1)$ vectors in $\mathcal
A$, such that $\mathbf a$ can be expressed by the convex combination
of these vectors. Hence there exists $\mathcal U_1^* \subset
\mathcal U_1$ satisfying $|\mathcal U_1^*|=k+1$, and coefficients
$\{\alpha_{u_1}\}_{u_1 \in \mathcal U_1^*}$ satisfying $\sum
\alpha_{u_1} = 1$ such that
\begin{eqnarray*}
p_{X_1}(x_1) &=& \sum_{u_1 \in \mathcal U_1^*} \alpha_{u_1}
f_{x_1}(u_1), \forall x_1 \in \mathcal X_1,\\
I(X_1;U_1) &=& \sum_{u_1 \in \mathcal U_1^*} \alpha_{u_1}
f_{R_1}(u_1),\\
\cdots \\
E[d_3(X_3,\widehat X^3)] &=& \sum_{u_1 \in \mathcal U_1^*}
\alpha_{u_1} f_{d_3}(u_1).
\end{eqnarray*}
Replacing $U_1$ by a new random variable $U_1^*$ on the alphabet
$\mathcal U_1^*$ with $Pr(U_1^*=u_1)=\alpha_{u_1}$, fixing the
conditional distributions $ p_{X_1|U_1^*}=p_{X_1|U_1}, p_{U_2^*|X^2
U_1^*}=p_{U_2|X^2 U_1}, p_{\widehat X_2^{3*}|X^3 U^{2*}}=p_{\widehat
X_2^3|X^3 U^2}$ and the function $g_1$, we preserve the marginal
distribution of $X_1$, all the mutual informations and expected
distortions in equations (\ref{eqn:CardRthm3_1}) -
(\ref{eqn:CardDthm3_3}). The progress is the new random variable
$U_1^*$ takes value in a smaller alphabet $\mathcal U_1^*$ whose
size is independent of $n$.

Note that because of the statistical structure of the joint
distribution
\[ p_{X^3U_1^{2*}\hat X_2^{3*}} = p_{U_1^*} p_{X_1|U_1^*} p_{X_2 X_3|X_1}
 p_{U_2^*|X^2 U_1^*} p_{\hat X_2^{3*}|X^3 U^{2*}},\]
the Markov chains $U_1^*-X_1 -X_2^3$ and $U_2^*-(X^2,U_1^*) -X_3$
hold. Because the marginal distribution $p_{X_1}$ is not changed,
the joint distribution $p_{X^3}$ also remains unchanged and
consisting with the requirement of the problem. However, the
distribution of $U_2$ and $\widehat X_3$ is possibly changed. So we
used $U_2^*$ and $\widehat X_3^*$ to indicate the corresponding
random variables associated with $U_1^*$. They still take values in
alphabets $\mathcal U_2$ and $\widehat{\mathcal X}^3$. The function
$g_1$ is unchanged, which means that $g_1(u_1,u_2) =
g_1(u_1^*,u_2^*)$ as long as $(u_1,u_2)=(u_1^*,u_2^*)$. But the
domain of $g_1$ shrinks from $\mathcal U_1\times \mathcal U_2$ to
$\mathcal U_1^*\times \mathcal U_2$.

Till now the alphabet $\mathcal U_1$ is reduced to $\mathcal U_1^*$
whose cardinality
\[|\mathcal U_1^*|=k+1=|\mathcal X_1|+6\]
is independent of $n$, while all the rate and distortion constraints
and the Markov chains still hold. Then we start to deal with the
alphabet $\mathcal U_2$.

Similar to the equations (\ref{eqn:CardXthm3_1}) to
(\ref{eqn:CardDthm3_3}), we can define functionals $f_{x_1 x_2
u_1^*}$ for all $(x_1, x_2, u_1^*) \in \mathcal X_1 \times \mathcal
X_2 \times \mathcal U_1^*$ and $f_{R_2'},
f_{R_3'},f_{d_1'},f_{d_2'}, f_{d_3'}$, such that
\begin{eqnarray}\label{eqn:CardXthm3_1prime}
p_{X^2 U_1^*}(x_1,x_2,u_1^*) &=:& \sum_{u_2 \in \mathcal U_2}
p_{U_2^*}(u_2) f_{x_1 x_2
u_1^*}(u_2,p_{X^2U_1^*|U_2^*}),\nonumber \\
&& \forall (x_1, x_2, u_1^*) \in \mathcal X_1 \times \mathcal X_2
\times \mathcal U_1^*,
\end{eqnarray}
\begin{equation}\label{eqn:CardRthm3_2prime}
I(X^2;U_2^*|U_1^*)  =:  \sum_{u_2 \in \mathcal U_2^*}
 p_{U_2^*}(u_2) f_{R_2'}(u_2,p_{X^2U_1^*|U_2^*}),
\end{equation}
\begin{equation}\label{eqn:CardRthm3_3prime}
I(X^3;\widehat X_3^*|U^{2*})
  =:  \sum_{u_2 \in \mathcal U_2^*} p_{U_2^*}(u_2)
   f_{R_3'}(u_2,p_{X^3U_1^*\hat X_3^*|U_2^*}),
\end{equation}
\begin{equation}\label{eqn:CardDthm3_1prime}
E[d_2(X_1,\widehat X_1^*)] =: \sum_{u_2 \in \mathcal U_2^*}
p_{U_2^*}(u_2) f_{d_1'}(u_2,p_{X_1U_1^*|U_2^*},g_1),
\end{equation}
\begin{equation}\label{eqn:CardDthm3_2prime}
E[d_2(X_2,\widehat X^{2*})] =:  \sum_{u_2 \in \mathcal U_2^*}
p_{U_2^*}(u_2) f_{d_2'}(u_2,p_{X_2 U_1^* \widehat X_2^*|U_2^*},g_1),
\end{equation}
\begin{equation}\label{eqn:CardDthm3_3prime}
E[d_3(X_3,\widehat X^{3*})] =:  \sum_{u_2 \in \mathcal U_2^*}
p_{U_2^*}(u_2) f_{d_3'}(u_2,p_{X_3 U_1^*\widehat
X_2^{3*}|U_2^*},g_1).
\end{equation}

Because the Markov chain $U_2^*-(X^2,U_1^*)-X_3$ holds, the joint
distribution can be written as follows,
\[ p_{X^3U_1^{2*}\widehat X_2^3} = p_{U_2^*} p_{X^2 U_1^*|U_2^*}
p_{X_3|X^2 U_1^*} p_{\widehat X_2^{3*}|X^3 U^{2*}}.\] Fixing $p_{X^2
U_1^*|U_2^*}, p_{X_3|X^2 U_1^*}, p_{\widehat X_2^{3*}|X^3 U^{2*}}$
and $g_1$, the functionals become functions depending solely on
$u_2$. Then following the same method, we can replace $U_2^*$ by
$U_2^{**}$ and preserve the marginal distribution $p_{X_1X_2U_1^*}$,
the mutual informations and expected distortions in equations
(\ref{eqn:CardRthm3_2prime})-(\ref{eqn:CardDthm3_3prime}). Because
altogether $|\mathcal X_1||\mathcal X_2||\mathcal U_1^*| + 4$
quantities should be preserved, one can limit the cardinality of
alphabet by
\[|\mathcal U_2^{**}|=|\mathcal X_1||\mathcal X_2||\mathcal
U_1^*| + 5=|\mathcal X_1|^2|\mathcal X_2|+6 |\mathcal X_1||\mathcal
X_2|+ 5.\]

In addition, by the statistical structure of the joint distribution,
the Markov chain $U_2^{**}-(X^2,U_1^*)-X_3$ can be verified to hold.
Finally, because the values of $(U_1^*, U_2^{**})$ never influence
the performance of the system, we can relabel them by
$\{1,2,\ldots,|\mathcal U_1^*|\} \times \{1,2,\ldots,|\mathcal
U_2^{**}|\}$ such that their values do not depend on the original
large size alphabets $\mathcal U_1, \mathcal U_2$. We completely
discard the old auxiliary random variables and rename the new random
variables $\{U_1^*,U_2^{**}\}$ by $\{U_1,U_2\}$ to continue the
proof.

Up to now we showed for any admissible rate-distortion tuple
$(\mathbf R, \mathbf D)$, $\forall \epsilon >0$, for all $n>
N(\epsilon)$, we can find $(U^2,\widehat X^3, g_1)$ satisfying
\begin{eqnarray}
&&R_1+\epsilon \geq I(X_1;U_1),\label{eqn:closurethm3_1}\\
&&R_2+\epsilon \geq I(X^2;U_2|U_1),\label{eqn:closurethm3_2}\\
&&R_3+\epsilon \geq I(X^3;\widehat X_2^3|U^2),\label{eqn:closurethm3_3}\\
&&D_j+\epsilon \geq E[d_j(X_j, \widehat X^j)], \ \ j=1,\ldots,T,\label{eqn:closurethm3_4}\\
&& \widehat X_1 =g_1(U^2),\nonumber \\
&& |\mathcal U_1|=|\mathcal X_1|+6,\nonumber \\
&& |\mathcal U_2|=|\mathcal X_1|^2|\mathcal X_2|+6 |\mathcal
X_1||\mathcal X_2|+ 5,\nonumber
\end{eqnarray}
and the Markov chains $U_1 - X_1 - X_2^3$ and $U_2 - (X^2,U_1) -
X_3$ hold, which implies
\begin{eqnarray}
&&I(U_1;X_2^3|X_1)=0,\label{eqn:closurethm3_5}\\
&&I(U_2;X_3|X^2,U_1)=0.\label{eqn:closurethm3_6}
\end{eqnarray}

\subsection{Taking limits}\label{appsec:closure}
Note that for each $(\epsilon, n)$, $|\mathcal U_j|$ is finite and
independent of $(\epsilon, n)$ for $j=1,2$. Therefore the
conditional distribution $p_{U^2,\widehat
X_2^3|X^3,\epsilon,n}(u^2,\hat x_2^3|x^3)$ is a finite dimensional
stochastic matrix taking values in a compact set, and
$g_{1,\epsilon,n}$ has only a finite number of possibilities.

Let $\{\epsilon_l\}_{l=1}^{\infty}$ be any sequence of real numbers
such that $\epsilon_l>0$ and $\epsilon_l \rightarrow 0$ as
$l\rightarrow \infty$. Let $\{n_l\}$ be any sequence of blocklengths
where $\forall l, n_l > N(\epsilon_l)$. Since $g_{1,\epsilon,n}$
takes values in a finite set, $\exists~g_1^*$ such that there exists
a subsequence $\{\epsilon_{l_i}\}_{i=1}^\infty$ such that for each
$\epsilon$ in this subsequence, $g_{1,\epsilon,n} \equiv g_1^*$.

Since $p_{U^2,\widehat X_2^3|X^3,\epsilon,n}$ lives in a compact
set, there exists again a subsequence of $\{p_{U^2,\widehat
X_2^3|X^3,\epsilon_{l_i},n_{l_i}}\}$ which converges to a limit
$p_{U^{*2},\widehat X_2^{*3}|X^3}$. Denote the auxiliary random
variables derived from the limit distribution by $(U^{*2},\widehat
X_2^{*3})$. Due to the continuity of conditional mutual information
and expectation with respect to probability distributions,
(\ref{eqn:closurethm3_1}) - (\ref{eqn:closurethm3_6}) become
\begin{eqnarray*}
&& R_1 \geq I(X_1;U_1^*), \\
&& R_2 \geq I(X^2;U_2^*|U_1^*), \\
&& R_3 \geq I(X^3;\widehat X_2^{*3}|U^{*2}), \\
&& D_j \geq E[d_j(X_j,\widehat{X}^{*j})], \ \
j=1,2,3,\\
&& I(U_1^*;X_2^3|X_1)=0,\\
&& I(U_2^*;X_3|X^2,U_1^*)=0,
\end{eqnarray*}
where $\widehat X_1^*:=g_1^*(U^{*2})$. The last two equalities imply
that the Markov chains $U_1^* - X_1 - X_2^3$ and $U_2^* -
(X^2,U_1^*) - X_3$ hold. Therefore $(\mathbf R,\mathbf D)$ belongs
to the right hand side of
(\ref{eqn:CNCrateregion}).\hspace*{\fill}~\QED

\footnotesize
\bibliography{mybibfile}

\end{document}